\documentclass[12pt]{iopart}

\usepackage[english]{babel}
\usepackage[utf8]{inputenc}
\usepackage{cite}

\expandafter\let\csname equation*\endcsname\relax
\expandafter\let\csname endequation*\endcsname\relax
\usepackage{amsmath}
\usepackage{mathtools}
\usepackage{amssymb}
\usepackage{amsfonts}

\usepackage{braket}
\usepackage{esint}
\usepackage{mathrsfs}
\usepackage{bbm}

\usepackage{graphicx, subfig}
\usepackage{sidecap}
\usepackage{psfrag}

\newcommand{\dd}{\text{d}}
\newcommand{\spp}{\mathrm{sp}}
\newcommand{\te}[1]{\mathrm{#1}}

\begin{document}

\title{A geometrical approach to the mean density of states in many-body quantum systems}

\author{Quirin Hummel\(^1\), Juan Diego Urbina and Klaus Richter}
\address{Institut f\"ur Theoretische Physik, Universit\"at Regensburg, D-93040 Regensburg, Germany}

\ead{\(^1\)quirin.hummel@physik.uni-regensburg.de}

\begin{abstract}
We present a novel analytical approach for the calculation of the mean density of states in many-body systems made of
confined indistinguishable and non-interacting particles.
Our method makes explicit the intrinsic geometry inherent in the symmetrization postulate and, in the spirit of the usual
Weyl expansion for the smooth part of the density of states in single-particle confined systems, our results take the form
of a sum over {\it clusters} of particles moving freely around manifolds in configuration space invariant under elements
of the group of permutations.
Being asymptotic, our approximation gives increasingly better results for large excitation energies and we
formally confirm that it coincides with the celebrated Bethe estimate in the appropriate region.
Moreover, our construction gives the correct high energy asymptotics expected from general considerations, and shows that
the emergence of the fermionic ground state is actually a consequence of an extremely delicate large cancellation effect.
Remarkably, our expansion in cluster zones is naturally incorporated for systems of interacting particles, opening the road
to address the fundamental problem about the interplay between confinement and interactions in many-body systems of identical particles.

\end{abstract}

\pacs{74.20.Fg, 75.10.Jm, 71.10.Li, 73.21.La}

\maketitle

\section{Introduction}

Understanding the general physical properties of interacting systems is one of the ultimate goals of classical and quantum mechanics,
and probably the most difficult.
An impressive toolbox including techniques such as the renormalization group~\cite{renom1,renom2}, perturbation expansions
in Feynman diagrams~\cite{pert1,pert2} and Density Functional Theory~\cite{DFT1,DFT2} has been developed during the past decades with
the sole objective of constructing the energy spectrum of quantum systems where external potentials, inter-particle interactions
and quantum statistics must be simultaneously considered.

In a non-relativistic first-quantization scenario (which is an effective low-energy approximation to the fundamental field-theoretical
relativistic description) both the Hilbert space and the Hamiltonian are fairly well understood
and the problem of constructing the energy spectrum is reduced to the,
still formidable, problem of calculating the many-body density of states
\begin{equation}
	\varrho_{\pm}(E,N)=-\frac{1}{\pi}\Im\left[\tr_{\pm}^{(N)}\hat{G}(E+i\epsilon)\right] \, .
\end{equation}
Here \( N \) is the total number of particles, \( (\pm) \) refers to the symmetry with respect to particle exchange of the subspace
where the trace operation is performed (fully symmetric wavefunction for bosons and fully antisymmetric one for fermions),
and \( \hat{G} \) is the Green function of the many-body system, defined in terms of the full interacting Hamiltonian
\begin{equation}
	\hat{H}=\hat{H}_{0}+g\hat{U}
\end{equation}
as
\begin{equation}
	\hat{G}(z)=\left(z-\hat{H}\right)^{-1} \, ,
\end{equation}
where \( \hat{H}_{0} \) describes a set of non-interacting particles, $\hat{U}$ is the (two-body) interaction potential and $g$
the coupling strength.

In this paper we are interested in the situation where the quantum system is made up of well defined microscopic degrees of
freedom either bosonic or fermionic, subject to a hard wall confinement in \( D \) dimensions such that the system is bounded
for all energies.
In this case, the density of states has the generic form
\begin{equation}
	\varrho_{\pm}(E,N)=\sum_{n}\delta(E-E_{n}^{(N)}) \, ,
\end{equation}
where \( n=1,2 \ldots \) labels the ordered eigenvalues \( E_{1}^{(N)},E_{2}^{(N)},\ldots \) of \( \hat{H} \) for fixed total
number of particles \( N \).
Although the precise knowledge of the density of states provides all the information about the spectrum of the system,
general considerations show that \( \varrho_{\pm}(E,N) \) can be unambiguously decomposed into a
smooth and an oscillatory part,
\begin{equation}
	\varrho_{\pm}(E,N)=\bar{\varrho}_{\pm}(E,N)+\varrho_{\pm}^\text{osc}(E,N) \, ,
\end{equation}
where the scaling of the smooth part is given asymptotically by the Weyl formula~\cite{Weyl,Baltes,Bohigas} 
\begin{equation}
	\bar{\varrho}_{\pm}(E,N)\sim 1 \slash \hbar^{DN} \, .
\end{equation}
If the interaction and the confinement is such that no constants of motion remain besides the total energy
and the classical many-body dynamics is chaotic,
for the oscillatory part we can formally apply the Gutzwiller trace formula~\cite{Gutzwiller,Brack} to get
\begin{equation}
	\label{eq rho osc prop}
	\varrho^\text{osc}_{\pm}(E,N)\sim 1 \slash \hbar
\end{equation}
in the semiclassical limit \( \hbar \rightarrow 0 \)
\footnote{We do not discuss questions~\cite{Primack} concerning the derivation of~(\ref{eq rho osc prop}) for \( D N \geq 3 \) here.}.
From the scaling with \( \hbar \) in the semiclassical limit it is expected that oscillatory contributions to the many-body
density of states in the strongly interacting regime where the quantum mechanical description of the system lies beyond the
single-particle picture, are extremely small compared to the mean density \( \bar{\varrho}_{\pm}(E,N) \).
Understandably, continuous effort has been dedicated to develop methods to construct this function either numerically or analytically
\cite{Bethe,Brack,pert2,nuclear2,nuclear3}.

To date, the only general way to obtain the precise \( \bar{\varrho}_{\pm}(E,N) \) is by explicit (typically numerical)
diagonalization techniques of the many-body problem followed by a convolution with a smoothing function \( W_{\epsilon}(E) \)
of width \( \epsilon \),
\begin{equation}
	\label{eq:barrho}
	\bar{\varrho}_{\pm}(E,N)=\int_{-\infty}^{\infty}\varrho_{\pm}(E,N)W_{\epsilon}(E)dE \, .
\end{equation}
However, due to the fast (exponential) growth of the basis required to achieve good convergence of the numerical results,
such numerical approaches can deal with only moderate numbers of particles
(see,~e.g.,~\cite{diag} for state of the art calculations).

Given the complexity of the problem, alternative methods are mandatory. Self-consistent mean-field methods, firmly grounded in the
Kohn-Sham theorem~\cite{DFT1,DFT2}, provide the most efficient way to
construct a set of single-particle wavefunctions such that the {\it ground state} energy of the interacting system can be systematically
approximated by artificial single-particle energies supplemented with the appropriate symmetry of the many-body wavefunction.

Despite the extremely successful application of self-consistent methods, ranging from nuclear physics to molecular systems, reaching chemical problems~\cite{DFT1,DFT2,DFT3}, the calculation of the many-body density of states within mean-field approaches faces a conceptually deep and basically unsolved problem. It is related with its very definition, and stems from the fact that the calculation of different types of observables actually requires a different definition for the mean field. Examples of this ambiguity are the calculation of ground-state vs excited-state energies and the construction of static vs dynamical properties of the system (for the calculation of transition amplitudes, for example, the mean field is necessarily time-dependent and depends on the initial and final states as well)~\cite{pert2}.

To be more precise, extending self-consistent approaches like the Density Functional Theory or Hartree-Fock method to excited states above the ground state requires a specific knowledge of which single-particle orbitals are optimized through the self-consistent equations, and therefore each excited many-body state requires a separated calculation leading to a different mean-field. Similar problems appear in formulations of the many-body problem based on a functional representation of the propagator where the mean-field is defined by means of a saddle-point equation: roughly speaking, the single-particle artificial potential used to mimic the effects of interparticle interactions becomes dependent on the excitation energy itself~\cite{pert2}.

Working with a mean-field which is simply a function of the position, independent on both excitation energy and/or time is then a strong assumption that lies behind much of the efforts to understand many-body systems in terms of a single-particle picture, an assumption which is in fact turned into essential in most cases where even a mild dependence of the mean field on the excitation energy would make the calculations impossible.

Luckily, many important physical effects accessible to experimental observation take place near or at the ground state energy, and it is expected that in this situation the mean field, providing the independent particle picture for the ground state within any of the self-consistent methods, can be used to calculate physical properties of excited states with decreasing accuracy as we move to higher and higher excitation energies. This physical consideration, together with the more pragmatic reasons explained above, has shown to be remarkably useful. In fact, the idea of an unique mean field allows us to use all the standard machinery of single-particle physics as input for the statistical mechanics results for independent particles and to finally produce experimentally accessible predictions beyond the strictly non-interacting case. A good mean field is then an excellent starting point.

A paradigm of the success of this approach is the study of low energy excitations of bounded fermionic systems with many particles like nuclei, metallic clusters and quantum dots~\cite{DFT1,DFT2,DFT3}. Here, a self-consistent calculation of variational type is set up to fix once and for all a single-particle potential responsible to replace the interparticle interaction. The result of the numerical calculation is used then to fit the parameters of a large family of functional forms (e.g.~Woods-Saxon for nuclei~\cite{WoodsSaxon1,WoodsSaxon2,WoodsSaxon3}) to finally produce an analytical form. Once this is achieved, standard methods of statistical physics are applied.

A result of these studies which is of importance for the present work is that, for a large enough number of electrons interacting through Coulomb forces in a billiard-shaped quantum dot, the mean field (in the sense of Hartree-Fock) closely follows the billiard potential. This allows us to assume a billiard model for the interacting system in mean field approximation knowing that it already captures part of the physics beyond the non-interacting case through a softening of the boundaries.

The program outlined above has reached a high sophistication, in particular when the single-particle physics is treated analytically by means of semiclassical methods, well suited to study the effective single-particle problem around the Fermi energy for the regime of large numbers of particles. The idea here is to construct the single-particle spectrum in terms of periodic orbits of the classical system considered as a single-particle moving within the mean field potential. The amount of work on the subject is huge and we refer to~\cite{Brack} for a more complete exposition including experimentally relevant effects like the existence of shell effects visible in the many-body density of states and due to continuous families of classical periodic orbits.

Another possibility offered by the mean field approach comes from one of the main lessons we have learnt from single-particle systems, namely that contrary to \( \varrho_{\pm}(E,N) \) which would be sensitive to every single detail of the microscopic {\it mean field} Hamiltonian, the smooth part of the
density of states \( \bar{\varrho}_{\pm}(E,N) \) predominantly depends on few {\it classical} quantities related to the measure of
the classical phase space manifolds at given energy.
Therefore, instead of going through numerically expensive calculations in order to construct the single-particle energies in
mean field approximation just to average out again the density of states and get its smooth part as indicated by~(\ref{eq:barrho}), one can try to construct and understand the classical phase-space structures behind . This approach can be carried on in two different and complementary ways.

The first option is to push the mean field picture and calculate the smooth part of the single-particle density of states using the Thomas-Fermi approximation and its variants~\cite{Brack}, and use it together with well established statistical techniques to construct \( \bar{\varrho}_{\pm}(E,N) \). This is the line of thought that lies behind the original attempt of Bethe~\cite{Bethe} to calculate the smooth part of the level density for nuclei and, more recently, a similar concept is followed in the seminal works of Weidenm\"uller~\cite{Weidenmueller} in order to formally construct the exact level density \( \varrho_\pm(E,N) \) for non-interacting particles. It is important to mention that this approach is based on the classical {\it single-particle} phase space, and the construction of the many-body density of states is purely formal and has no direct interpretation in terms of the classical {\it many-body} phase space.

The second possibility, namely, to calculate the smooth part of the many-body density of states in mean field approximation by relating it directly to the structure of the many-body classical phase space has not been systematically addressed before. In our opinion, such approach has the potential advantage of avoiding the {\it a priori} conflict with the inclusion of residual interactions inherent to any approach that is based on the single-particle phase space.

As we will show, these two approaches give quite different points of view for the calculation of \( \bar{\varrho}_{\pm}(E,N) \), and their equivalence for the strict mean field limit is by no means trivial. However, once this equivalence is established, our methods based on the many-body phase space will allow to address the fundamental question concerning how the relevant geometrical structures are translated into the problem in the presence of residual interactions. This is the program we propose here.

Once the single-particle picture is adopted the study of reference is the so-called Bethe estimate, providing an asymptotic result
for the density of states in many-fermion systems in mean field approximation, valid for energies far enough
(in units of the single-particle mean level spacing) from the ground state \( E_\text{GS} \) and large numbers of particles~\cite{Bethe},
\begin{equation}
	\label{eq:bethe}
	\bar{\varrho}_{\pm}(E,N)\simeq \frac{{\rm e}^{\sqrt{\frac{2 \pi^{2}}{3} \bar{\varrho}_\spp(E_\te{F}) (E-E_\text{GS})}}}
	{\sqrt{48}(E-E_\text{GS})} \, .
\end{equation}
The essential aspect of Bethe's result in the context of interest here is that it is an asymptotic approximation for the smooth
part of the density of states, and can be interpreted as the fermionic analogue of the Weyl expansion for billiard systems.
However, and despite its enormous importance,~(\ref{eq:bethe}) (and the thermodynamic formalism used in its derivation)
is of limited use for us, as it does not provide any clue on how classical phase space manifolds work together to produce both
its characteristic functional form and the scale \( E_\text{GS} \).

In the physical scenario of interest here (the application of Bethe's method in confined electronic systems) the most relevant
problems besides the all long issue of residual interactions are, i) the inclusion of finite
\( N \) effects, ii) the consistent treatment of the ground state energy and the density of states {\it around} it, and iii)
the emergence of the standard Weyl expansion \( \bar{\varrho}_{\pm}(E,N) \sim E^{DN/2-1} \) for high enough energies.
These questions have been addressed already, and the amount of literature in the subject is extensive, so we will only briefly
review the state of the art.

Finite size effects on the asymptotic density of states in systems of fermions can be systematically calculated in mean field
approximation by extending Bethe's result, which is the leading order in an expansion valid for large excitation energies and
large numbers of particles obtained by a saddle point approximation of the exact grand-canonical partition function
\cite{Bethe,pert2,nuclear2,nuclear3}.
Because the problem of counting many-body eigenstates for non-interacting identical particles has a natural combinatorial formulation,
this approach has produced an unexpected and fruitful interaction with number theory.
In this spirit, corrections to the Bethe estimate arising from finite number of particles, oscillatory corrections to the single-particle
density of states, and shell effects affecting the ground state energy have been considered~\cite{LeboeufFinite}.

The status of the ground state energy within the asymptotic approach is somehow delicate, as strictly speaking, the many-body density
of states must vanish identically for energies bellow \( E_\text{GS} \). As obvious as this observation may be, it turns out that it is extremely
difficult to construct a theory providing the large energy asymptotics for \( \bar{\varrho}_{\pm}(E,N) \) while keeping this condition exact.
This problem may be considered at first glance a merely academic, since after all no physical process takes place for energies
bellow the ground state, and the later has a very precise definition in terms of the single-particle density of states.
We must however keep in mind that the final goal of any approach is to deal with the fully interacting system and/or to provide
insight and better methods to define and calculate the mean field and the effect of the residual interactions.
Beyond the mean field picture, we eventually need a systematic method to identify and construct the ground state energy
independent of the counting prescription valid in the non-interacting case.
To the best of our knowledge, a systematic study of how do approximations for \( \bar{\varrho}_{\pm}(E,N) \) behave for
\( E < E_\text{GS} \) is missing.

Finally, very general and robust considerations demand that asymptotically (when the energy goes to infinity), essential quantum
mechanical effects such as the non-zero ground state energy gradually disappear and a purely classical description emerges.
In this limit, one should recover the standard Weyl expansion for the density of states where quantum symmetry effects only appear
as a global reduction of the available phase space to a fraction of \( 1 \slash N! \) due to the identity of the particles but not to their particular statistics~\cite{Berry}.
Using asymptotic methods to understand this transition leads to some interesting results connected with the number-theoretical
formulation of the problem~\cite{LeboeufFinite}.

Together with the goal of providing a geometrical approach to \( \bar{\varrho}_{\pm}(E,N) \), the last paragraphs
indicate the main motivations of the present work.
We attempt to provide a method to construct the Weyl approximation to the smooth part of the density of states which relies
only on kinematic and geometrical aspects within the mean field approximation.
As expected, our results are connected with several others (in particular with Bethe's) and an important aspect of our work is
to make these connections explicit.
However, the method itself and the physical idea behind it, namely that the Weyl expansion can be systematically constructed
out of free propagation near symmetry manifolds are our novel contributions to the subject.

The paper is organised as follows.
In section~\ref{sec non-int dist prtcls} we introduce the basic notation and briefly review the construction of the Weyl
expansion for systems of non-interacting identical but distinguishable particles.
In sections~\ref{non-int undist prtcls} and~\ref{sec equiv of MB and SP} the symmetrization postulate is used to give the formal expression
for the full density of states for indistinguishable particles, and we show how this formal object can be understood in
terms of the geometry of a higher dimensional phase space when the fundamental domain associated with the group of permutations \( S_N \) is considered.
The role of classical manifolds invariant under different elements of the symmetric group is clearly seen using the example
of two fermions on a line in section~\ref{sec 2 f on a line}.
The very non-trivial generalization of this construction for arbitrary type of particles (bosons or fermions),
arbitrary dimension \( D \) of the single-particle configuration space \( \Omega \) and arbitrary number of particles \( N \) is fully carried out
in sections~\ref{sec gen case} and~\ref{sec Weyl exp of non-int prtcls} and culminate with a full identification of the classical manifolds and their measures responsible for the functional form of \( \bar{\varrho}_{\pm}(E,N) \) and the emergence of \( E_\text{GS} \).
We analyse our results in sections~\ref{sec geom emergence} and~\ref{sec comp w known results}, where the equivalence of our results with the Weidenm\"uller convolution formula and with Bethe's estimate are rigorously proved.
We conclude and discuss the extension of our results for the interacting case in section~\ref{sec outlook}.

\section{Non-interacting Distinguishable Particles}
\label{sec non-int dist prtcls}%
	First consider a billiard system of \( N \) non-interacting, identical, but distinguishable particles in \( D \) dimensions,
	specified by their coordinates
	\begin{align}
		{\bf q}_i \in \mathbb{R}^D \, , \qquad i=1,\ldots,N
	\end{align}
	with spatial components
	\begin{align}
		q_i^{(d)} \in \mathbb{R} \, , \qquad d=1,\ldots,D \, .
	\end{align}
	This corresponds to an effective \( N \cdot D \)-dimensional billiard system of a
	single particle described by
	\begin{align}
		\label{eq q def ND}
		{\bf q} = ( {\bf q}_1, \ldots , {\bf q}_N ) \in \mathbb{R}^{ND} \, .
	\end{align}
	In this case the only effect of particle exchange symmetry is thereby the
	existence of discrete spatial symmetries in the effective higher dimensional single-particle system. But due to distinguishability,
	there is no restriction to any subset of wavefunctions with specific symmetry under exchange transformation.

	In general, the density of states (DOS) of a bound, time independent system can be expressed as the inverse Laplace transform of the trace
	of the propagator \mbox{\( \hat{U}(t) = \exp ( -i \slash \hbar \hat{H} t ) \)}:
	\begin{align}
		\label{eq DOS laplace}
		\varrho(E) = \mathscr{L}^{-1}_\beta \left[ \int \dd^{ND}q \; K({\bf q},{\bf q}; t=-i \hbar \beta) \right](E) \, ,
	\end{align}
	where the trace is performed in coordinate space with \( K({\bf q}',{\bf q};t) = \bra{{\bf q}'} \hat{U}(t) \ket{\bf q} \) and the
	inverse Laplace transform has to be applied with respect to the variable \( \beta \).

	In semiclassical approximation (corresponding to the formal limit \( \hbar \rightarrow 0 \))
	there are two contributions to the DOS
	\begin{align}
		\varrho^\te{scl}(E) = \varrho^\te{osc}(E) + \bar{\varrho}(E) \, ,
	\end{align}
	one oscillating (\( \varrho^\te{osc} \)) and one smooth (\( \bar{\varrho} \)) in the energy \( E \).
	The oscillatory part arises from various stationary phase approximations in~(\ref{eq DOS laplace}) starting from a path integral representation
	of the propagator. The process leads to a description by periodic orbits of the underlying classical system. The oscillatory part of
	the DOS is then in semiclassical approximation expressed by the Gutzwiller trace formula in the chaotic case~\cite{Gutzwiller,BerryMount}
	and the Berry-Tabor trace formula in the integrable case~\cite{BerryTabor} respectively.

	The smooth part of the DOS is related to short path contributions that are not caught by periodic orbits in the analysis of the
	trace of the propagator.
	Reflecting the short time behaviour of the propagator, these are related to the assumption of local free quantum propagation and
	additional boundary corrections~\cite{BalianBloch}. The Weyl expansion for the \( N \cdot D \)-dimensional billiard reads~\cite{BalianBloch}
	\begin{align}
	\label{eq Weyl ND}
	\begin{split}
		\bar{\varrho}(E) =  &\left( \frac{m}{2 \pi \hbar^2} \right)^\frac{ND}{2} \frac{V_{ND}}{\Gamma\left( \frac{ND}{2} \right)}
			E^{\frac{ND}{2}-1} \theta(E) \\
		&\pm \frac{1}{4} \left( \frac{m}{2 \pi \hbar^2} \right)^\frac{ND-1}{2} \frac{S_{ND-1}}{\Gamma\left( \frac{ND-1}{2} \right)}
			E^{\frac{ND-1}{2}-1} \theta(E) + \cdots \, .
	\end{split}
	\end{align}
	The first term in~(\ref{eq Weyl ND}) will be referred to as the volume Weyl term and equals the Thomas-Fermi approximation
	\( \bar{\varrho}_\te{TF}(E) \)~\cite{Brack} proportional to the \( N \cdot D \)-dimensional volume \( V_{ND} \).
	The second term originates from wave reflection near billiard boundaries under the assumption of local flatness of the surface.
	This involves free quantum propagation to mirror points yielding a fast converging integral over the coordinate perpendicular to the surface
	in~(\ref{eq DOS laplace}). Thus the second term is proportional to the integral over parallel coordinates yielding the surface
	\( S_{ND-1} \) (not to be confused with the symmetric group \( S_N \)) of the billiard instead of its volume.
	Higher corrections in the expansion correspond to propagation between multiply reflected image points accounting for curvature,
	edges or corners of the boundary.
\section{Non-interacting Undistinguishable Particles}
\label{non-int undist prtcls}%
	In the case of \( N \) identical particles that are undistinguishable, the state of the system obeys a specific symmetry with respect
	to particle exchange. The state is either symmetric or antisymmetric under the exchange of any two particles, depending on whether
	they are bosons or fermions.
	\begin{align}
		\hat{P} \lvert \psi_\pm \rangle &= (\pm 1)^P \; \lvert \psi_\pm \rangle \, , \qquad (-1)^P \coloneqq \mathrm{sgn}(P)
	\end{align}
	for any permutation \( P \in S_N \), where \( \hat{P} \) is the corresponding permutation operator acting on many-body states,
	and plus and minus refer to bosons respectively fermions.
	In order to obtain the physical spectrum of such a system one has to restrict to those eigenenergies that
	correspond to the subspace of Hilbert space with appropriate symmetry. Let \( \hat{\mathbbm{1}}_\pm = \hat{\mathbbm{1}}_\pm^\dagger\)
	be the projector onto the subspace of correct symmetry.
	Restricting the trace in~(\ref{eq DOS laplace})
	to this subspace is equivalent to replacing the propagator by its symmetry projected analogue
	\begin{align}
		\label{eq sym proj prop U}
		\hat{U}_\pm(t) &\coloneqq \hat{\mathbbm{1}}_\pm \hat{U}(t) \hat{\mathbbm{1}}_\pm = \hat{\mathbbm{1}}_\pm \hat{U}(t) \, , \\
		\label{eq sym proj prop}
		K_\pm({\bf q}', {\bf q}; t) &\coloneqq \frac{1}{N!} \sum_{P \in S_N} (\pm 1)^P \; K(P{\bf q}', {\bf q}; t) \, .
	\end{align}
	In~(\ref{eq sym proj prop U}), the commutation of the time evolution operator and the symmetry projector due
	to \( [ \hat{P}, \hat{H} ] = 0 \) and idempotence of \( \hat{\mathbbm{1}}_\pm \) have been used.
	This leads to the symmetry projected DOS
	\begin{align}
		\label{eq DOS sym laplace}
		\varrho_\pm(E) = \mathscr{L}^{-1}_\beta \left[ \frac{1}{N!} \sum_{P \in S_N} (\pm 1)^P
			\int \dd^{N D}q \, K(P{\bf q}, {\bf q}; t = - i \hbar \beta) \right](E) \, .
	\end{align}
	Thus symmetry causes the need of taking wave propagation over finite distance into account, as \( P{\bf q} \neq {\bf q} \) in general.
	For the oscillating part in \( \varrho_\pm(E) \) it is possible to construct a fundamental domain in phase space where it
	is again sufficient to find periodic orbits and additionally the group characters \( (\pm 1)^P \) of the group elements \( P \)
	connected to each trajectory \( ({\bf q},{\bf p}) \mapsto (P{\bf q},P{\bf p}) \) in the unfolded full domain~\cite{Robbins}.
	However this procedure has no direct application to the smooth part of the DOS in the general case
	of arbitrary particle number and spatial dimension.

	This is true even in a bosonic system in \( D > 1 \) despite the possibility of mapping it to the higher dimensional system of a
	single particle without symmetry moving in the fundamental domain in coordinate space with topological identification of symmetry
	related points. That is because of the non-trivial structure of such a wrapped fundamental domain especially in the vicinity
	of symmetry planes defined by \( P{\bf q} = {\bf q} \) for some \( P \in S_N \). In order to illustrate this, compare to a two
	dimensional single-particle system with discrete rotational symmetry
	\begin{align}
		\begin{split}
			[ \hat{H}, \hat{R}_\phi ] &= 0 \, , \\
			\hat{R}_\phi &= \exp \Bigl( -\frac{i}{\hbar} \hat{L}_z \phi \Bigr) \, , \\
			\phi &= \frac{2 \pi}{n} \, , \qquad n \in \mathbb{N} \, .
		\end{split}
	\end{align}
	The restriction to the subspace of states symmetric under the elementary rotation
	\( \hat{R}_\phi \ket{\psi_\te{sym}} = \ket{\psi_\te{sym}} \)
	is equivalent to the restriction to the wrapped fundamental domain with usual wave dynamics. The wrapped fundamental domain
	in this example is the restriction of the billiard to a sector of central angle \( \phi \) with identification of points along
	its two bordering half-lines. As it is equal to a cone, this produces non-trivial wave propagation at the origin
	which gives rise to additional corrections in the level density \( \bar{\varrho}_\te{sym}(E) \) of symmetric states~\cite{LauritzenWhelan}.
	Analogue to that, mapping a bosonic system to its wrapped fundamental domain implies non-trivial wave propagation in the vicinity
	of the symmetry planes. This shows that it is reasonable to stay in the full domain for the calculation of the smooth part
	\( \bar{\varrho}_\pm(E) \), and this is the approach we will follow here.

	Previous to the treatment of the general case it is instructive to analyse the simple example of many identical fermions on a line which
	can be mapped to a fundamental domain where the additional correction due to symmetry can easily be obtained by usual methods.
\section{Equivalence of Many-Body and High Dimensional Single-Particle Pictures}
\label{sec equiv of MB and SP}%
	This section will mainly focus on systems of many fermions moving on a line of length \(L\).
	These systems have some special properties that are setting them apart from higher dimensional ones.
	It is worth restricting to such systems for a moment since they can easily be mapped
	onto single-particle systems.

\begin{figure}
		\begin{center}
			{\psfrag{q1}[tl]{\(q_1\)}%
			\psfrag{q2}[br]{\(q_2\)}%
			\psfrag{q3}[br]{\(q_3\)}%
			\psfrag{L1}[bl]{\(L\)}%
			\psfrag{L2}[tl]{\(L\)}%
			\psfrag{L3}[cl]{\(L\)}%
			\includegraphics[width=0.3\textwidth]{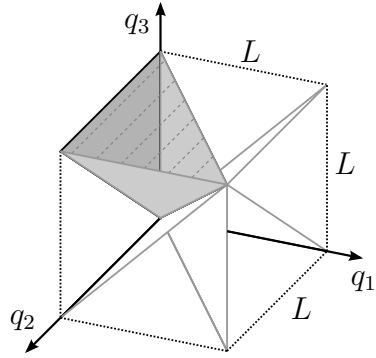}}%
      \caption{\label{fig: 3p1d fund dom}Fundamental domain \(\mathscr{F}\) for \(N=3, D=1\), bounded by symmetry planes
								\( q_1 = q_2 \) and \( q_2 = q_3 \) (light grey) and physical boundaries
								\( q_1 = 0 \) and \( q_3 = L \) (dark grey and shaded).}
		\end{center}
	\end{figure}

	In one dimensional systems a fundamental domain in coordinate space can be given by
	\begin{align}
		\mathscr{F} := \Bigl\{ {\bf q} \in \mathbb{R}^N \Big| 0 \leq q_1 \leq q_2 \leq \cdots \leq q_N \leq L \Bigr\} \, .
	\end{align}
	Its boundaries are given by the equations
	\begin{align}
		\label{eq sym planes}
		q_i &= q_{i+1} \, , \qquad i=1,\ldots,N-1
	\end{align}
	due to symmetry related reduction and
	\begin{align}
		\begin{split}
			\label{eq phys conf 1d}
			q_1 &= 0 \, , \\
			q_N &= L
		\end{split}
	\end{align}
	for the physical confinement to the line \( [0,L] \).
	An example of this construction in the case of three fermions is shown in figure~\ref{fig: 3p1d fund dom}.
	The full domain can be reobtained by applying
	all possible permutations \( P \in S_N \) to the fundamental domain
	\begin{equation}
		[0,L]^N = \bigcup_{P \in S_N} P \left( \mathscr{F} \right) \, .
	\end{equation}
	Due to \(D=1\), topological identification of boundary points in this context is not needed.

	For fermions the restriction to antisymmetric states yields the condition of vanishing wave function all along the boundary
	\begin{equation}
		\label{eq dirichlet}
		\psi \left( {\bf q} \right) = 0 \qquad (\forall {\bf q}) (\exists i \neq j) (q_i = q_j) \, .
	\end{equation}
	As a Dirichlet boundary condition, this condition is sufficient to determine the eigenfunctions in \( \mathscr{F} \)
	together with the single-particle conditions~(\ref{eq phys conf 1d}).
	The symmetry planes~(\ref{eq sym planes}) can be thought of as hard walls or in other words infinite potential barriers.
	The values of the wave functions in the other parts of the full domain are then obtained by
	\begin{equation}
		\label{eq wave function from F to full}
		\psi \left( P \, {\bf q} \right) = (-1)^P \, \psi \left( {\bf q} \right) \, .
	\end{equation}
	Thus a \(1D\) billiard with \( N \) fermions is equivalent to a single-particle billiard of dimension \( N \cdot D \),
	in which the usual Weyl expansion can be used to obtain \( \bar{\varrho}_-(E) \).

	One has to stress that this is a feature of one dimensional systems only because there no additional condition besides~(\ref{eq dirichlet})
	is imposed on the wave function within the fundamental domain.
	In contrast to that, the corresponding condition \( \psi({\bf q}) = 0 \) 
	for dimensions larger than one is not given along the whole boundaries of the fundamental domain,
	but instead only on lower dimensional manifolds embedded in those.
	The reason is that for \( D>1 \), \( \mathscr{F} \) is not bounded by the symmetry planes \( {\bf q}_i = {\bf q}_j \),
	which have a dimension of \( N \cdot D - D < N \cdot D - 1 \) and therefore are not able to separate volumes in
	\( N \cdot D \)-dimensional configuration space.
	Rather the restriction to one of the spatial components in the conditions that define the symmetry planes are
	able to do so. One choice of fundamental domain is
	\begin{align}
		\mathscr{F} := \Bigl\{ {\bf q} \in \Omega^N \Big| q_1^{(1)} \leq q_2^{(1)} \leq \cdots \leq q_N^{(1)} \Bigr\} \, ,
	\end{align}
	where its symmetry related boundaries are defined through
	\begin{align}
		q_i^{(1)} = q_{i+1}^{(1)} \, , \qquad i=1,\ldots,N-1 \, ,
	\end{align}
	and \( {\bf q} \in \Omega^N \) denotes the physical confinement to the interior \( \Omega \subset \mathbb{R}^D \) of the billiard.
	The symmetry planes \( {\bf q}_i = {\bf q}_{i+1} \), on which~(\ref{eq wave function from F to full}) imposes vanishing of the
	wave function, are only lower dimensional submanifolds embedded in the boundaries. Thus, the
	condition~(\ref{eq wave function from F to full}) can not be expressed as a Dirichlet boundary condition in a fundamental domain.

	Moreover, even in the case of \(D=1\) sharp edges in the boundary
	of \( \mathscr{F} \) can cause problems for \( N > 2 \), making the expansion of Balian and Bloch~\cite{BalianBloch}
	inapplicable, which is a generalization of the Weyl expansion to arbitrary dimensions but smooth boundaries.%
	\footnote{It is worth to note that one could topologically identify points along the boundary of \( \mathscr{F} \) that are related
	by symmetry and thereby create a fundamental domain with complex topology. We will call this object the {\it wrapped fundamental domain}.
	The problem in the fermionic case then would be the loss of continuity of the wavefunction because of sign inversion
	\( \psi \left( {\bf q} \right) \rightarrow - \psi \left( {\bf q} \right) \) in the direction
	perpendicular to boundaries related to odd permutations.
	This condition seems quite peculiar and so far the author has not found a treatment of it in~\cite{BalianBloch}.}
\section{Two Non-Interacting Fermions on a Line}
\label{sec 2 f on a line}%
\begin{figure}
		\begin{center}
		\hspace*{\fill}
		{\psfrag{q1}{\(q_1\)}%
		\psfrag{q2}{\(q_2\)}%
		\psfrag{id}{id}%
		\psfrag{P}{\(P\)}%
		\subfloat[2 particles in 1 D]{\label{fig: 2p1d bla}\includegraphics[width=0.38\textwidth]{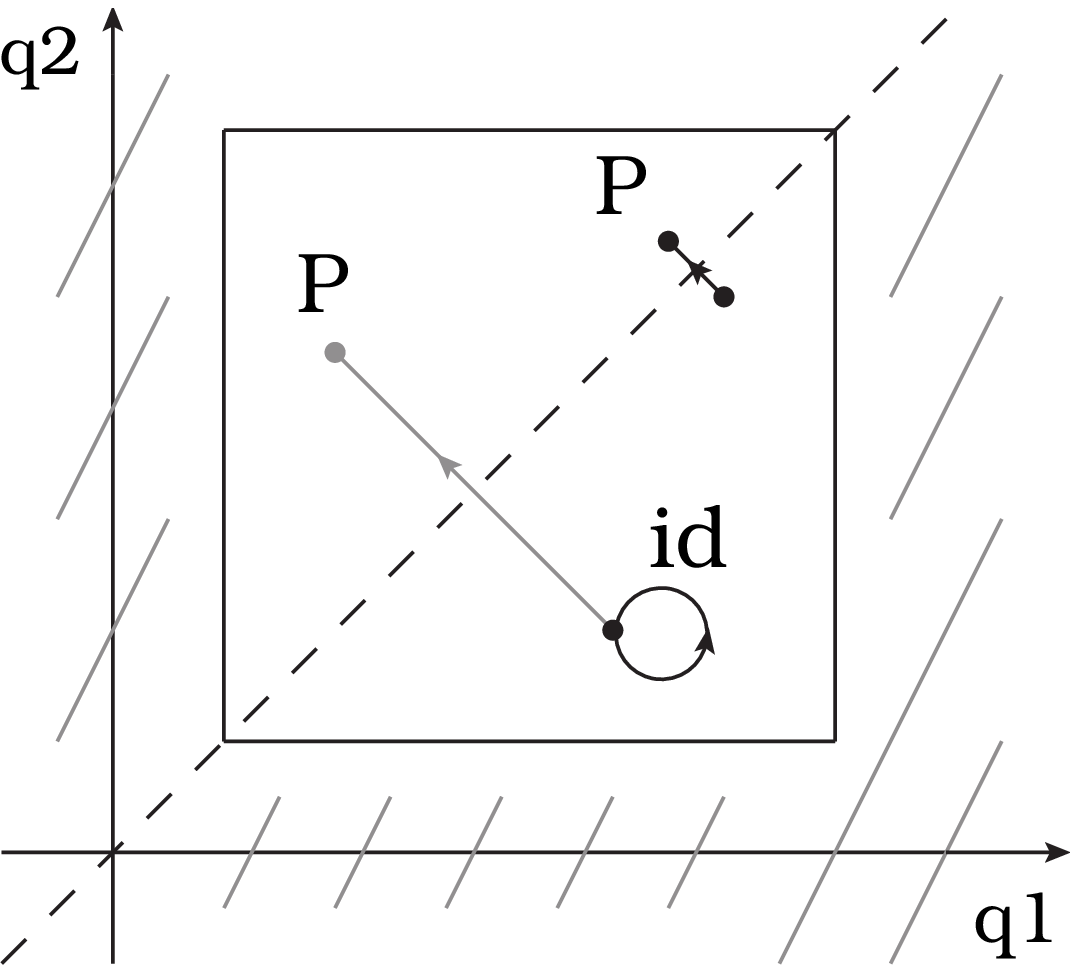}}}%
		\hspace*{\fill}
		{\psfrag{V0}{\(V=0\)}%
		\psfrag{Vinf}{\(V=\infty\)}%
		\subfloat[1 particle in 2 D]{\label{fig: 1p2d bla}\includegraphics[width=0.38\textwidth]{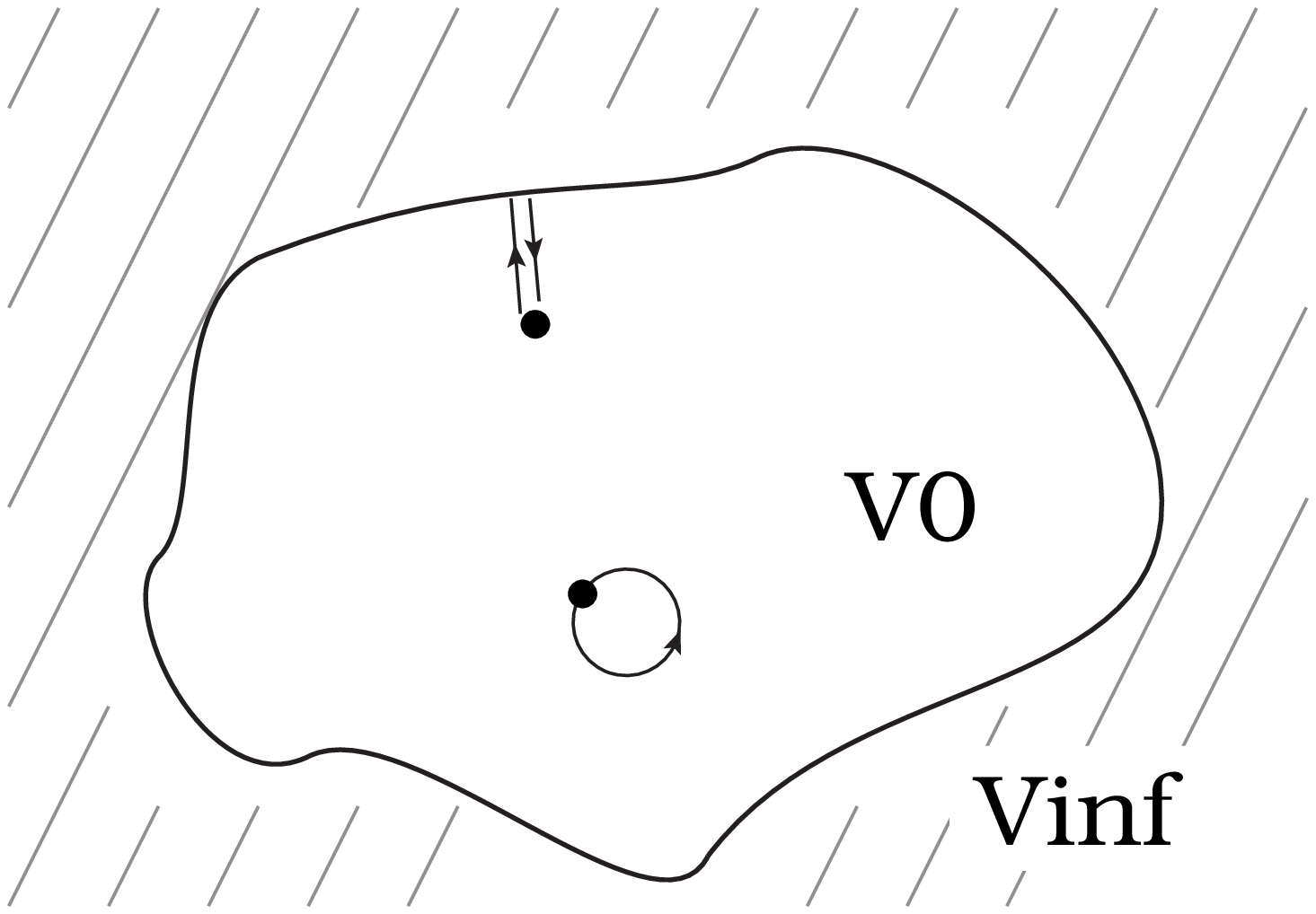}}}%
		\hspace*{\fill}
		\caption{\label{fig: 2p1d 1p2d} Comparison of the system of two identical particles in one dimension and one particle in
				two dimensions.}
		\end{center}
	\end{figure}

	Consider a one-dimensional system of two fermions confined to a line of length \( L \).
	The only two permutations are the identity and their exchange.
	Figure~\ref{fig: 2p1d bla} illustrates the two possible contributions to the propagator.
	Since the system has an effective two-dimensional description it is straightforward to compare it
	to a single-particle two-dimensional billiard. There, \( \bar{\varrho}(E) \) is made up of contributions
	from free propagation and from reflections on the boundary, which is illustrated in figure~\ref{fig: 1p2d bla}.

\begin{figure}
		\begin{center}
		{\psfrag{q1}{\(q_1\)}%
		\psfrag{q2}{\(q_2\)}%
		\psfrag{K01234}[tc]{\(\qquad \quad \; \; K(P{\bf q}, {\bf q}; t)\)}%
		\subfloat[2 particles in 1 D]{\label{fig: 2p1d blubb}\includegraphics[width=0.38\textwidth]{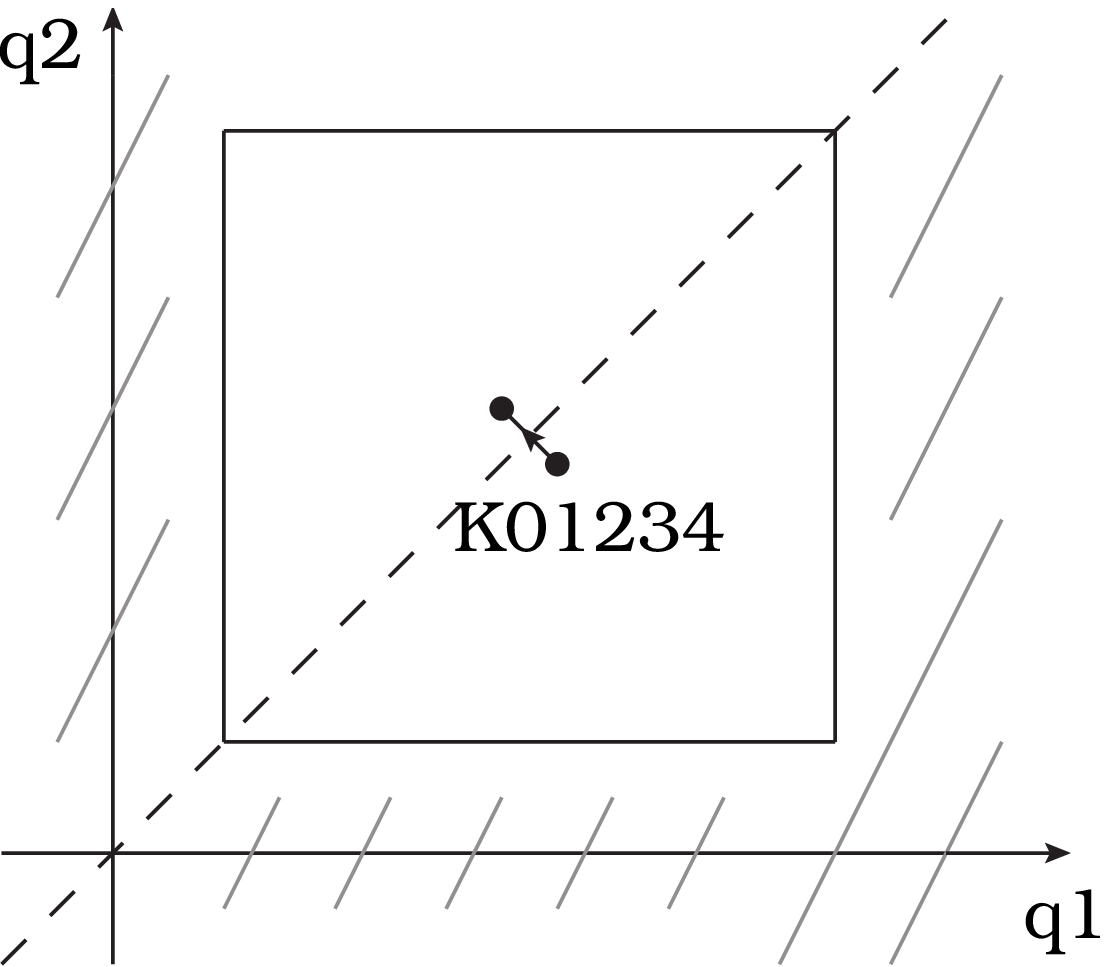}}}%
		\hfil
		{\psfrag{q1}{\(q_1\)}%
		\psfrag{q2}{\(q_2\)}%
		\psfrag{K01234}[cl]{\(K(R{\bf q}, {\bf q}; t)\)}%
		\subfloat[1 particle in 2 D]{\label{fig: 1p2d blubb}\includegraphics[width=0.38\textwidth]{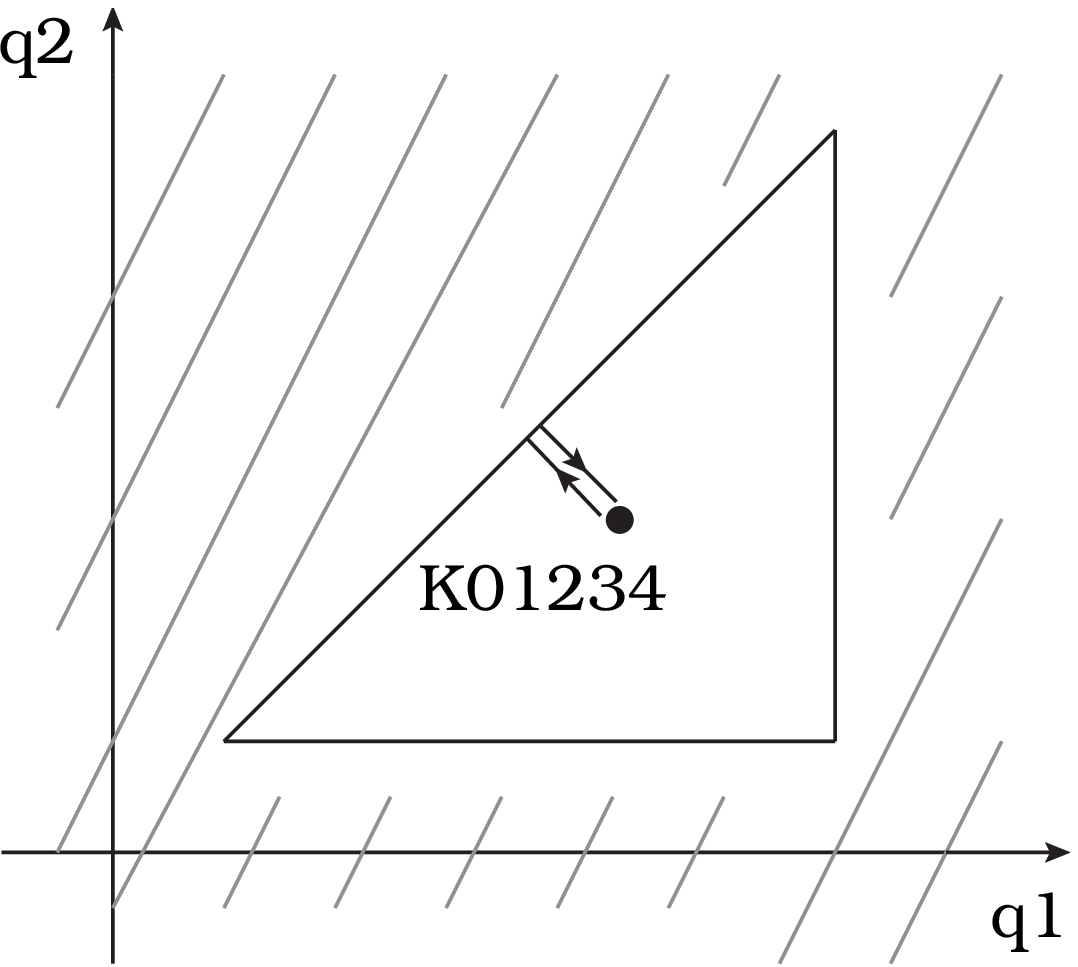}}}%
		\caption{ \label{2p1d blubb} Equivalence of the system of two identical particles in one dimension and one particle in the
				two-dimensional fundamental domain.}
		\end{center}
	\end{figure}

	As discussed in section~\ref{sec equiv of MB and SP} there is a simple two-dimensional single
	particle billiard that is exactly equivalent to the two particle system. That is, the billiard defined by the fundamental domain,
	here chosen as \( \mathscr{F} : L \geq q_1 \geq q_2 \geq 0 \) with an additional hard wall boundary along the symmetry line
	\( q_1 = q_2 \). In this two-dimensional
	picture reflections on the additional boundary, which are addressed by propagation to mirror points, are mapped to the propagation
	with respect to the exchange permutation in the one dimensional
	many-body picture. Figure~\ref{2p1d blubb} illustrates the corresponding propagation in both pictures.
	The smooth part of the DOS of the two-fermion system including corner corrections reads
	\begin{align}
		\label{eq rho 2p1d}
		\bar{\varrho}_-(E)= &\frac{L^2}{8 \pi} \left( \frac{2m}{\hbar^2} \right) \, \theta(E) \, - \, (2 + \sqrt{2})
		\frac{L}{8 \pi} \left( \frac{2m}{\hbar^2} \right)^{\frac{1}{2}} E^{-\frac{1}{2}} \theta(E) \,+\, \frac{3}{8} \delta(E) \, .
	\end{align}
	The Weyl expansion~(\ref{eq rho 2p1d}) includes a volume term \( \bar{\varrho}_\te{v}(E) \) with area \mbox{\( A = L^2 \slash 2 \)},
	where the factor of \( \frac{1}{2} \) originating in the restriction to the fundamental domain corresponds to the
	factor \( 1 \slash N! \) in the symmetry projected propagator~(\ref{eq sym proj prop}).
	Thus this factor in the volume term, which corresponds to taking into account only the identity permutation,
	is the leading order effect of exchange symmetry.
	In general exchange corrections related to all other permutations are of sub-leading order as they correspond to Weyl-like boundary
	corrections. For example, in the case at hand, the only exchange permutation yields in leading order the perimeter correction proportional
	to \( \sqrt{2} \).
	Nevertheless, in general all exchange corrections are important to give physically reasonable results, as will be shown in the next
	sections.
\section{General Case - Propagation in Cluster Zones}
\label{sec gen case}%
\subsection{Invariant Manifolds and Cluster Zones}
\begin{figure}
		\begin{center}
			\includegraphics[width=0.41\textwidth]{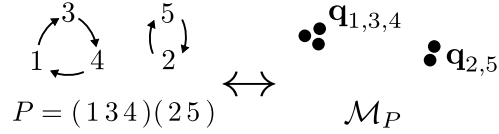}
			\caption{\label{fig: cycles manifold}Example of the correspondence between the cycle decomposition of a permutation \( P \) and the
				invariant manifold \( \mathcal{M}_P \) pictured as the associated clustering of particles.}
		\end{center}
	\end{figure}

	The last section showed that in the example of two particles on a line the correction to \( \bar{\varrho}_-(E) \) due to the exchange
	permutation is related to the propagation in the vicinity of the symmetry line \( q_1 = q_2 \) just as the inclusion of wave
	reflections in a single-particle billiard only affects the short time propagation near the physical boundary. The symmetry
	line is characterised by the invariance under \( P = (\,1\,2\,) \), so that the distance \( | P{\bf q} - {\bf q} | \) becomes zero.
	This is the very reason to assume short path contributions to come from its vicinity. The concept of invariant manifolds
	is extended to the general case by finding the manifolds associated with each permutation \( P \), defined by
	\begin{align}
		\mathcal{M}_P = \bigl\{ {\bf q} \in \mathbb{R}^{ND} \; \big| \; | P{\bf q} - {\bf q} | = 0 \bigr\} \cap \Omega^N \, .
	\end{align}
	\( P \) can be written as a composition of commuting cycles (see for example~\cite{cycledecomp})
	\begin{align}
		\label{eq P cycle decomposition}
		&P = \sigma_1 \cdots \sigma_l \, ,
	\intertext{acting on distinct sets of particle indexes of size}
		\label{eq P cycle decomposition end}
		&N_1, N_2, \ldots , N_l \, .
	\end{align}
	So we see that \( \mathcal{M}_P \) is the manifold defined by the coincidence of the coordinates of all particles associated with each cycle
	\begin{align}
		\label{eq def of MP}
		\mathcal{M}_P = \bigcap\limits_{\omega=1}^l \bigl\{ {\bf q} \in \Omega^N \; \big| \;
			{\bf q}_i = {\bf q}_j \quad \forall i,j \in I_\omega \bigr\} \, .
	\end{align}
	As a simple example take the permutation
	\begin{align}
		P = (\,1\,3\,4\,)\,(\,2\,5\,) \, ,
	\end{align}
	whose associated manifold \( \mathcal{M}_P \) corresponds to the condition (see figure~\ref{fig: cycles manifold})
	\begin{align}
		\bigl( {\bf q}_1 = {\bf q}_3 = {\bf q}_4 \bigr) \, \wedge \, \bigl( {\bf q}_2 = {\bf q}_5 \bigr) \, .
	\end{align}%
	In contrast to the symmetry line in the one-dimensional two particle case, these manifolds in general can not be seen as a boundary
	or surface in coordinate space in the sense of dividing the space into distinct pieces.
	The vicinities of these invariant manifolds
	shall from now on be referred to as {\it cluster zones}.
	All particles associated with a particular cycle index subset will be subsumed to the notion of a {\it cluster}.
	A system that is momentarily arranged in a particular cluster zone is composed of \( l \) clusters, each associated
	with a cycle in \( P \). Each cluster \( \omega \) is composed of \( N_\omega \) particles according to the length of the cycle.

	A separation into coordinates parallel to \( \mathcal{M}_P \) and perpendicular
	to it suggests itself, since the propagation near \( \mathcal{M}_P \) will not depend on shifting the position along them.
	This holds at least as long as one does not get too close to the single-particle billiard boundaries
	to be included later. First, these single-particle boundary corrections shall be
	neglected, assuming the propagation to be invariant along \( \mathcal{M}_P \). This will be referred to as the {\it unconfined case}.

	Furthermore, the invariance of propagation along the invariant manifolds is in a strict treatment also broken in the case of interaction.
	But the intention lying behind this construction is that when restricting to interactions of rather short range, the propagation
	can be assumed to be invariant along \( \mathcal{M}_P \) as long as one does not get too close to other invariant submanifolds.
	In other words, as long as the coordinates corresponding to different cycles do not become too close. Or to put
	it into a more intuitive picture, the different clusters should not collide.
	A discussion on that can be found in the last section.
\subsection{The Measure of Invariant Manifolds}
	The infinitesimal volume element \( \dd \mu \) on \( \mathcal{M}_P \) is determined by the infinitesimal
	vectors in full \( (ND) \)-dimensional coordinate space lying in \( \mathcal{M}_P \) that correspond to the variations of
	independent coordinates.

	First, for each permutation \( P \) in cycle decomposition~(\ref{eq P cycle decomposition}--\ref{eq P cycle decomposition end})
	the particles are relabelled without loss of generality in a way that the first cycle \( \sigma_1 \)
	involves the first \( N_1 \) particles, the second cycle \( \sigma_2 \) involves the \( N_2 \) subsequent particles, and so on.

	Since all particles associated with one cycle have to fulfil the condition of equal coordinates
	(\ref{eq def of MP}), there is exactly one independent \( D \)-dimensional vector for each cycle
	\( \sigma_\omega \, , \, \omega=1,\ldots,l \) ,
	while all other particles in the same cycle have to follow in order to remain on \( \mathcal{M}_P \).
	Let the \( l \) independent vectors be denoted by
	\begin{align}
		{\bf x}_\omega = ( x_\omega^{(1)}, \ldots, x_\omega^{(D)} ) \in \Omega \, , \qquad \omega=1,\ldots,l \, .
	\end{align}
	With these preliminaries, any point on \( \mathcal{M}_P \) is described by
	\begin{align}
		\label{eq q from x}
		{\bf q} = ( \underbrace{{\bf x}_1, \ldots, {\bf x}_1}_{N_1}, \underbrace{{\bf x}_2, \ldots, {\bf x}_2}_{N_2} ,
				\ldots, \underbrace{{\bf x}_l, \ldots, {\bf x}_l}_{N_l} ) \in \Omega^N \, .
	\end{align}
	\\
	The infinitesimal volume element \( \dd \mu \) of \( \mathcal{M}_P \) is the volume of the parallelotope spanned by all
	\( l \cdot D \) infinitesimal tangent vectors
	\begin{align}
		\frac{\partial {\bf q}}{\partial x_\omega^{(d)}} \; \dd x_\omega^{(d)} \, , \qquad \omega=1,\ldots,l , \quad d=1,\ldots,D \, .
	\end{align}
	\( ( \dd \mu )^2 \) equals the Gramian determinant \( G \) of all these vectors, which is the determinant of the matrix
	made up of all pairwise scalar products.
	From~(\ref{eq q from x}) one obtains
	\begin{align}
		\frac{\partial {\bf q}}{\partial x_\omega^{(d)}} &=
				( \underbrace{{\bf 0}, \ldots \ldots, {\bf 0}}_{\mathclap{N_1 + \cdots + N_{\omega-1}}},
				\underbrace{{\bf e}_d , \ldots , {\bf e}_d}_{N_\omega} ,
				\underbrace{{\bf 0} , \ldots \ldots, {\bf 0}}_{N_{\omega+1} + \cdots + N_l} ) \\
	\intertext{with}
		{\bf e}_d &= ( 0, \ldots, 0, \underset{\mathclap{\substack{\uparrow \\ d\te{-th}}}}{ 1} , 0 , \ldots, 0 ) \in \mathbb{R}^D
	\end{align}
	and therefore
	\begin{align}
		\left< \frac{\partial {\bf q}}{\partial x_\omega^{(d)}} , \frac{\partial {\bf q}}{\partial x_{\omega'}^{(d')}} \right>
				&= \delta_{d d'} \delta_{\omega \omega'} N_{\omega} \, .
	\end{align}
	The Gramian determinant reads
	\begin{align}
		G = \det \Bigl( \text{diag}( \underbrace{N_1, \ldots, N_1}_{D} , \underbrace{N_2, \ldots, N_2}_{D} , \ldots,
				\underbrace{N_l, \ldots, N_l}_{D} ) \Bigr) \cdot \Bigl( \prod_{\omega=1}^l \dd^D x_\omega \Bigr)^2 \, ,
	\end{align}
	\begin{align}
		\dd \mu = \sqrt{G} =  \sqrt{N_1}^D \cdots \sqrt{N_l}^D \; \cdot \; \dd^D x_1 \, \cdots \, \dd^D x_l \, .
	\end{align}
	The total measure of the manifold \( \mu(\mathcal{M}_P) \) is obtained by integration of all independent coordinates
	\( {\bf x}_\omega \) over \( \Omega \):
	\begin{align}
		\label{eq manifold measure}
		\mu(\mathcal{M}_P) &= \int \dd \mu = V_D^l \, \cdot \, \Bigl( \prod_{\omega=1}^l N_\omega \Bigr)^\frac{D}{2} \, , \\
	\intertext{with the \( D \)-dimensional volume of the billiard}
		V_D &= \int\limits_\Omega \dd^D x_\omega \, .
	\end{align}
	\\
	This measure is only depending on the partition of \( N \) into integers \( N_1 + \cdots + N_l \) corresponding to the decomposition
	of \( P \) into cycles with lengths \( N_1 , \ldots , N_l \). Each permutation associated with the particular partition
	\( \{ N_1 , \ldots , N_l \} \) yields an invariant manifold of the same measure~(\ref{eq manifold measure}).
	Furthermore the contributions from short path propagation in their vicinities will be the same due to the symmetry with respect to 
	relabelling particle indexes. This allows the replacement of the sum over all permutations by a sum over all distinct partitions
	of \( N \) with an additional factor of
	\begin{align}
		\label{eq comb factor}
		c(N_1,\ldots,N_l) = N! \; \Bigl( \prod_{\omega=1}^l \frac{1}{N_\omega} \Bigr) \, \Bigl( \prod_{n=1}^N \frac{1}{m_n!} \Bigr)
	\end{align}
	in each summand, which is the number of permutations \( P \in S_N \) with cycle lengths \( \{ N_1 , \ldots , N_l \} \)
	in their cycle decomposition.
	Thereby, \( m_n \) denotes the multiplicity the cycle length \( n \) appears with.
\section{The Weyl Expansion of Non-Interacting Particles}
\label{sec Weyl exp of non-int prtcls}%
	Up to this point, the analysis is quite general and is valid also for interacting systems.
	Although we believe it to be feasible in the context of interaction, for now the non-interacting case will be carried out explicitly.
	Furthermore, in this calculation, effects of the physical boundary are omitted.

	Under these assumptions the propagator in
	(\ref{eq sym proj prop}) is taken as the product of free propagators of all particles
	\begin{align}
		\label{eq K non-int}
		K({\bf q}^\prime, {\bf q}; t) = \prod_{i=1}^N K_0({\bf q}^\prime_i , {\bf q}_i ; t) \, .
	\end{align}
	At the end of this section, the full expression
	including effects of (locally flat) physical boundary can be found. The corresponding calculation is shown in the appendix.

	For the calculation of the summand corresponding to a particular permutation \( P \) in~(\ref{eq DOS sym laplace}), the particle indexes of
	different cycles in \( P \) don not mix up, yielding a product of independent propagators, which can be traced separately,
	each factor corresponding to a specific cycle in the decomposition~(\ref{eq P cycle decomposition}).
	Consider now the trace of all coordinates corresponding to the cycle \( \sigma_\omega \).
	For this purpose, the particles are relabelled, so that the indexes associated with \( \sigma_\omega \) are simply
	\( I_\omega = \{ 1,\ldots N_\omega \} \).
	For the sake of simplicity we write \( {\bf q} \) and \( P{\bf q} \) by meaning
	the restrictions to the first \( N_\omega \) particles
	\begin{align}
		\begin{split}
			\label{eq q def nD}
			{\bf q} &= ( {\bf q}_1 , {\bf q}_2, \ldots , {\bf q}_n ) \, , \\
			P{\bf q} &= ( {\bf q}_2 , \ldots , {\bf q}_n, {\bf q}_1 ) \, .
		\end{split}
	\end{align}
	Furthermore note the abbreviation \( n=N_\omega \). The integral over the associated co\-ordinates reads
	\begin{align}
		\label{eq trace free prop n D}
		\int \dd^{n D}q \, K_0(P{\bf q}, {\bf q}; t) = \int \dd^{n D}q \, \left( \frac{m}{2 \pi \hbar i t} \right)^\frac{nD}{2}
			\, \exp\Bigl({\frac{i}{\hbar} \, \frac{m}{2t} | P{\bf q} - {\bf q} |^2 }\Bigr) \, ,
	\end{align}
	where the equality of a product of \( n \) free propagators in \( D \) dimensions and one \( n \cdot D \)-dimensional
	free propagator has been used. With the distance vector
	\begin{align}
		P{\bf q}-{\bf q} &= ({\bf q}_2-{\bf q}_1,{\bf q}_3-{\bf q}_2,\ldots,{\bf q}_1 -{\bf q}_n) \, ,
	\intertext{the squared distance is}
		| P{\bf q} - {\bf q} |^2 &= | {\bf q}_2-{\bf q}_1 |^2 + \cdots + | {\bf q}_1-{\bf q}_n |^2 \nonumber \\
		\label{eq squared dist 1}
		&= \sum_{d=1}^D \bigl[ ( q^{(d)}_2-q^{(d)}_1 )^2 + \cdots + ( q^{(d)}_1-q^{(d)}_n )^2 \bigr] \, .
	\end{align}
	The overall squared distance is the sum of squared distances according to one spatial component, which are just the
	summands in~(\ref{eq squared dist 1}).
	The following calculation proceeds in equal manner for all spatial components \( d = 1,\ldots,D \).
	So the notation is further simplified by calculating only the factor corresponding to one spatial component in
	(\ref{eq trace free prop n D}) and by omitting the superscript \( (d) \).
	For this calculation in \( n \)-dimensional space we simply write \( {\bf q} \) and \( P{\bf q} \) by meaning
	the corresponding tuples of one particular spatial component.
	\begin{align}
		\begin{split}
			\label{eq q def n}
			{\bf q} &= ( q_1 , q_2 , q_3 , \ldots , q_n ) \, , \\
			P{\bf q} &= ( q_2 , q_3 ,\ldots , q_n , q_1 ) \, .
		\end{split}
	\end{align}
	Which of the definitions~(\ref{eq q def ND}), (\ref{eq q def nD}) and~(\ref{eq q def n}) is used in a particular
	step will be clear from the number of regarded dimensions in the context.

	In this simplified notation~(\ref{eq q def n}) each summand in~(\ref{eq squared dist 1}) is
	\begin{align}
		\label{eq squared dist 2}
		| P{\bf q} - {\bf q} |^2 = ( q_2-q_1 )^2 + \cdots + ( q_n-q_{n-1} )^2 + ( q_1-q_n )^2 \, .
	\end{align}
	The trace to calculate is
	\begin{align}
		\label{eq trace free prop n}
		\int \dd^nq \, K_0(P{\bf q}, {\bf q}; t) = \int \dd^nq \, \left( \frac{m}{2 \pi \hbar i t} \right)^\frac{n}{2}
			\, \exp\Bigl({\frac{i}{\hbar} \, \frac{m}{2t} | P{\bf q} - {\bf q} |^2 }\Bigr) \, .
	\end{align}
	The squared distance is of second order in all coordinates, enabling to perform the integral 
	(\ref{eq trace free prop n}) as a generalized multidimensional Gaussian integral
	\begin{align}
		\label{eq multi gauss 1}
		\int \dd^{m}x \; \exp\Bigl({-\frac{1}{2} {\bf x}^T M {\bf x}}\Bigr) = \sqrt{\frac{(2 \pi)^m}{\det(M)}}
			\, , \quad M = M^T \in \text{GL}_m \, ,
	\end{align}
	which will not be used directly, since the determinant of \( M \) equals zero. It has one eigenvalue \( \lambda=0 \) corresponding
	to the direction parallel to the invariant manifold, expressing the local translational invariance along
	\begin{align}
		\hat{\bf q}_\parallel &= \frac{1}{\sqrt{n}} (1,\ldots,1) \, ,
	\intertext{since the distance vector is invariant in this direction, }
		P({\bf q}+a\hat{\bf q}_\parallel) - ({\bf q}+a\hat{\bf q}_\parallel) &= P{\bf q}- {\bf q} \; + \;
			a( \underbrace{P \hat{\bf q}_\parallel - \hat{\bf q}_\parallel}_{=\,0} ) = P{\bf q}- {\bf q} \, .
	\end{align}
	This is the very reason why the separation into coordinates parallel and perpendicular to the invariant
	manifolds suggests itself.
	One way to proceed would be to introduce suitable perpendicular coordinates, perform the corresponding lower
	dimensional integral of the propagator and in the end multiply it by the measure of the invariant manifold.
	This would be the direct analogue to the usual computation of the surface correction in the single-particle
	Weyl expansion~(\ref{eq Weyl ND}).
	One has to stress that in the interacting case this is most likely the most convenient way to calculate
	the trace of the propagator.
	And indeed, one can follow this procedure in the non-interacting case.
	But the introduction of perpendicular coordinates is rather uncomfortable since naturally one is lead to non-orthogonal
	coordinate systems and therefore has to introduce a metric tensor and pay extra attention to the arising volume elements.
	Although the calculation for the free case can be carried out in this manner, we will present an alternative approach
	in the non-interacting case that is more convenient and straightforward.

	For the following analysis, a minimum cycle length of \( n \geq 2 \) is assumed.
	The trivial case \( n = 1 \) will be included automatically in the resulting expressions.
	In the \( n \)-dimensional space of particle coordinates corresponding to only one cycle and only one spatial dimension,
	the subspace of vectors under which the squared distance is invariant is only one-dimensional (there is only one
	\( \hat{\bf q}_\parallel \)). Accordingly, the matrix \( M \) has exactly one eigenvalue that is
	vanishing when bringing the trace~(\ref{eq trace free prop n}) into the form of~(\ref{eq multi gauss 1}).
	Therefore it is sufficient to separate one of the \( n \) coordinates, e.g.~\( q_1 \) and calculate
	the integral over all others as a generalized multidimensional Gaussian integral with linear term
	\begin{align}
		\label{eq multi gauss 2}
			\int \dd^{m}x \; \exp\Bigl({-\frac{1}{2} {\bf x}^T M {\bf x} \, + \, {\bf v}^T {\bf x}}\Bigr) =
				\sqrt{\frac{(2 \pi)^n}{\det(M)}} \; \exp\Bigl({\frac{1}{2} {\bf v}^T M^{-1} {\bf v}}\Bigr)
	\end{align}
	with \( {\bf v} \in \mathbb{C}^m \) and \( M = M^T \in \text{GL}_m \).
	The remaining integral \( \int \dd q_1 \) can then be kept and eventually, when considering all spatial components and cycles,
	it will automatically produce the measure of \( \mathcal{M}_P \) together with the determinant prefactors. Parts of the prefactors
	will thereby act as the Jacobian determinant associated with the relation of the volume element of the manifold to the independent
	coordinates \( q_{1,\omega}^{(d)} \) , \( d=1,\ldots,D \; ,\omega=1,\ldots,l \). We abbreviate
	\begin{align}
		\alpha =  \frac{i}{\hbar} \, \frac{m}{2t}
	\end{align}
	and write~(\ref{eq trace free prop n}) as
	\begin{align}
		\label{eq trace free prop n A B}
		\left( -\frac{\alpha}{\pi} \right)^\frac{n}{2}  \int \dd q_1 \exp\bigl({2 \alpha q_1^2 }\bigr)
			\int \dd q_2 \cdots \dd q_n
			\exp \Bigl( -\frac{\alpha}{2} \sum_{i,j=1}^{n-1} A_{ij} q_{i+1} q_{j+1}  +  \alpha \sum_{i=1}^{n-1} b_i q_{i+1} \Bigr)
	\end{align}
	with some symmetric matrix \( A \) and a vector \( {\bf b} \), which are identified by separating all \mbox{\( q_1 \)-}dependent
	terms in~(\ref{eq squared dist 2}):
	\begin{align}
		\begin{split}
			| P{\bf q} - {\bf q} |^2 &= 2 q_1^2 +
				\underbrace{\sum_{i=2}^n 2 q_i^2 - \sum_{i=2}^{n-1} 2 q_i q_{i+1}}_{= -\frac{1}{2} \sum A_{ij} q_{i+1} q_{j+1}} \;
				\underbrace{\vphantom{\sum_{i=2}^{n-1}}- 2 q_1 q_2 - 2 q_1 q_n}_{= \sum b_i q_{i+1}} \, , \\
			A_{ij} &= -4 \left[ \delta_{ij} - \frac{1}{2} ( \delta_{i, j+1} + \delta_{i+1,j} ) \right] \, , \\
			\label{eq b identify}
			b_i &= -2 q_1 ( \delta_{i1} + \delta_{i,n+1} ) \, , \qquad \qquad \qquad i,j=1,\ldots,n-1 \, .
		\end{split}
	\end{align}
	\( A \) is a tridiagonal matrix of dimension \( n-1 \)
	\begin{align}
		A &= (-4)\cdot \begin{pmatrix}
				1 & -\frac{1}{2} & 0 & 0 & \cdots \\
				-\frac{1}{2} & 1 & -\frac{1}{2} & 0 & \cdots \\
				0 & -\frac{1}{2} & 1 & -\frac{1}{2} & & \\
				0 & 0 & -\frac{1}{2} & \ddots & \ddots \\
				\vdots & \vdots & & \ddots & \ddots
									\end{pmatrix} \, ,
	\intertext{whose inverse and determinant are}
		\label{eq A inverse}
		(A^{-1})_{ij} &= \left\{ \begin{array}{ll}
			-\frac{1}{2} j (1 - \frac{i}{n}) \quad &i \geq j \\
			(A^{-1})_{ji} &i < j ,\\
														\end{array} \right. \, , \\
		\det(A) &= n (-2)^{n-1} \, .
	\end{align}
	Using~(\ref{eq b identify}) and~(\ref{eq A inverse}) gives
	\begin{align}
		\begin{split}
			{\bf b}^T A^{-1} {\bf b} &= 4 q_1^2 [ (A^{-1})_{11} + (A^{-1})_{1,n-1} + (A^{-1})_{n-1,1} + (A^{-1})_{n-1, n-1} ] \\
			&= 4 q_1^2 \Bigl( -\frac{1}{2n} \Bigr) [ n - 1 + 2 + n - 1 ] \\
			&= - 4 q_1^2 \, .
		\end{split}
	\end{align}
	With this and~(\ref{eq multi gauss 2}) the whole integral~(\ref{eq trace free prop n A B}) becomes
	\begin{align}
		\left( -\frac{\alpha}{\pi} \right)^\frac{n}{2}  \sqrt{\frac{(2 \pi)^{n-1}}{\alpha^{n-1}\det(A)}}
			\int \dd q_1 \, \exp\Bigl({2 \alpha q_1^2 } + {\frac{1}{2} \alpha {\bf b}^T A^{-1} {\bf b}}\Bigr)
			= \left( -\frac{\alpha}{\pi} \right)^\frac{1}{2}  n^{-\frac{1}{2}}  \int \dd q_1 \, .
	\end{align}
	\\
	By collecting all spatial components we get the contribution~(\ref{eq trace free prop n D}) corresponding to a particular cycle
	\begin{align}
		\int \dd^{n D}q \, K_0((P{\bf q}), {\bf q}; t) = \left( -\frac{\alpha}{\pi} \right)^\frac{D}{2}  n^{-\frac{D}{2}}
			\int\limits_\Omega \dd^D q_1 = \left( \frac{m}{2 \pi \hbar i t} \right)^\frac{D}{2}  N_\omega^{-\frac{D}{2}}  V_D \, ,
	\end{align}
	where we reintroduced the notations~(\ref{eq q def nD})
	and \( n = N_\omega \) for the length of the particular cycle under investigation. Note that this general form also includes the
	case of a one-cycle \( N_\omega = 1 \).

	By considering all traces corresponding to the cycles of one particular permutation one gets
	\begin{align}
		\label{eq trace free prop P ND}
		\int \dd^{N D}q \, K_0((P{\bf q}), {\bf q}; t) = \left( \frac{m}{2 \pi \hbar i t} \right)^\frac{l D}{2}
			\Bigl( \prod_{\omega=1}^l N_\omega \Bigr)^{-\frac{D}{2}}  V_D^l \, .
	\end{align}
	(\ref{eq trace free prop P ND}) as an expression associated with a permutation \( P \) only depends on the partition of \( N \)
	into cycle lengths (as does the measure \( \mu( \mathcal{M}_P ) \)~(\ref{eq manifold measure})).
	By collecting all permutations with the same partition in the sum over \( S_N \), the trace of the symmetry projected
	propagator can be written as
	\begin{align}
		\begin{split}
			\int \dd^{N D}q \, K_{0,\pm}({\bf q}, {\bf q}; t) = &\frac{1}{N!}
				\sum_{l=1}^N (\pm 1)^{N-l} \sum_{\mathclap{\substack{N_1,\ldots,N_l =1\\ N_1 \leq \cdots \leq N_l}}}^N
				\delta_{N, \, \sum N_\omega} \; c(N_1,\ldots,N_l)  \\
			& \times \; \; \left( \frac{m}{2 \pi \hbar i t} \right)^\frac{l D}{2} 
				\Bigl( \prod_{\omega=1}^l N_\omega \Bigr)^{-\frac{D}{2}}  V_D^l \, .
		\end{split}
	\end{align}
	As before, \( c(N_1,\ldots,N_l) \) denotes the number of permutations with a cycle decomposition of lengths \( N_1,\ldots,N_l \)
	(\ref{eq comb factor}).
	Using the bilateral Laplace transformation rule
	\begin{align}
		\mathcal{L}^{-1}_\tau \left[ \frac{\Gamma(\nu)}{\tau^\nu} \right](x)
			= \frac{1}{2 \pi i} \int_{\epsilon-i\infty}^{\epsilon+i\infty} \!\! \dd \tau \,
			\frac{\Gamma(\nu)}{\tau^\nu}\,{\rm e}^{\tau x} = x^{\nu-1} \theta(x) \, , \qquad \nu>0 \, ,
	\end{align}
	with the Heaviside step function \( \theta(x) \),
	the final result for the smooth part of the symmetry projected DOS~(\ref{eq DOS sym laplace})
	for a system of \( N \) identical non-interacting bosons (\(+\)) or fermions (\(-\)) in a \( D \)-dimensional billiard
	of volume \( V_D \) without confinement corrections reads
	\begin{align}
		\label{eq rho sym unconfined}
		\begin{split}
			\bar{\varrho}_\pm(E) = &\frac{1}{N!} \sum_{l=1}^N (\pm 1)^{N-l}
				\left[ \quad \; \sum_{\mathclap{\substack{N_1,\ldots,N_l =1\\ N_1 \leq \cdots \leq N_l}}}^N
				\delta_{N, \, \sum N_\omega} \; c(N_1,\ldots,N_l) \Bigl( \prod_{\omega=1}^l \frac{1}{N_\omega} \Bigr)^\frac{D}{2} \right] \\
			&\times \; \left( \frac{m}{2 \pi \hbar^2} \right)^\frac{l D}{2}
				  \frac{V_D^l}{\Gamma\left(\frac{l D}{2}\right)}
				E^{\frac{l D}{2}-1}  \theta(E) \, .
		\end{split}
	\end{align}
	In general~(\ref{eq rho sym unconfined}) is a sum of powers of \( E \) with coefficients that are,
	besides their dependence on the billiard volume, expressed as sums over partitions \( N = N_1 + \cdots + N_l \)
	only depending on \( N, l \) and \( D \) and therefore universal for all \( N \)-particle billiard systems with dimension \( D \).
	A more compact form of~(\ref{eq rho sym unconfined}) can be obtained by appropriately scaling the density
	and energy and rewriting the sum over partitions as ordered tuples.
	Using~(\ref{eq comb factor}) and defining the scaling-density
	\begin{align}
		\label{eq rho0}
		\varrho_0 &= \frac{m \sqrt[D]{V_D}^2}{2 \pi \hbar^2} \quad \Leftrightarrow \quad
			\bar{\varrho}_\spp(E) = \varrho_0 \frac{(\varrho_0 E)^{\frac{D}{2}-1}}{\Gamma\left(\frac{D}{2}\right)} \, ,
	\end{align}
		which is roughly the density at the first single-particle level, and the universal coefficients
	\begin{align}
		\label{eq univ coeff unconfined}
		C_{l} &= \sum_{{\substack{N_1,\ldots,N_l =1\\ \sum N_\omega = N}}}^N
			\Bigl( \prod_{\omega=1}^l \frac{1}{N_\omega} \Bigr)^{\frac{D}{2}+1} \, ,
	\end{align}
		leads to
	\begin{align}
		\label{eq rho sym unconfined 2}
		\begin{split}
			\bar{\varrho}_\pm(E) &= \varrho_0 \sum_{l=1}^N \frac{(\pm 1)^{N-l}}{l!}
				C_{l} \frac{ ( \varrho_0 E)^{\frac{l D}{2} - 1} }
				{\Gamma\left( \frac{l D}{2} \right)} \theta(E) \, .
		\end{split}
	\end{align}
	An extension of~(\ref{eq rho sym unconfined 2}) is given by the inclusion of physical boundary effects.
	Its derivation under the assumption of locally flat boundaries can be performed directly by incorporating
	the additional geometrical features to the propagation in cluster zones.
	But as this way to proceed gets extensive and seems a bit long-winded, an alternative, indirect but equivalent derivation
	utilizing a convolution formula~(\ref{eq weidenmuller conv}) by Weidenm\"uller~\cite{Weidenmueller} is chosen for
	this publication and can be found in \ref{app calc conv confined}.
	The resulting expression reads
	\begin{align}
		\label{eq rho sym confined}
			\bar{\varrho}_\pm(E) = \varrho_0 \sum_{l=1}^N (\pm 1)^{N-l}
				\sum_{\substack{l_\text{V}, l_\text{S}=0 \\ \sum l_i = l}}^l
				\frac{ C_{l,l_\text{V}} \gamma^{l_\text{S}} }{ l_\text{V}! \, l_\text{S}! } \,
				\frac{ ( \varrho_0 E)^{\lambda - 1} }{ \Gamma( \lambda ) } \theta(E) \, ,
	\end{align}
	where \( \lambda = l_\text{V} D \slash 2 + l_\text{S} (D-1) \slash 2 \).
	It depends on the universal coefficients
	\begin{align}
		\label{eq univ coeff confined}
		C_{l,l_\text{V}} = \sum_{{\substack{N_1,\ldots,N_l =1\\ \sum N_\omega = N}}}^N
			\Bigl( \prod_{\omega=1}^{l_\text{V}} \frac{1}{N_\omega} \Bigr)^{\frac{D}{2}+1}
			\Bigl( \prod_{\omega=l_\text{V}+1}^l \frac{1}{N_\omega} \Bigr)^{\frac{D}{2}+\frac{1}{2}} \, ,
	\end{align}
	and the dimensionless geometrical parameter
	\begin{align}
		\label{eq def gamma}
		\gamma = \pm \frac{S_{D-1}}{4 \sqrt[D]{V_D}^{D-1}} \, ,
	\end{align}
	which represents the ratio of the surface \( S_{D-1} \) to the volume \( V_D \) of the billiard.
	The plus(minus) sign in~(\ref{eq def gamma}) refers to van Neumann(Dirichlet) conditions at the physical boundary.
	Equations~(\ref{eq rho sym unconfined 2}) and~(\ref{eq rho sym confined}) can be regarded as the main results of this publication.
	\( l_\text{V} + l_\text{S} = l \) represents the splitting of all \( l \) clusters into such that contribute via
	free propagation in the interior (\( l_\text{V} \)) and such that contribute by reflection along the boundary (\( l_\text{S} \))
	of the billiard. In the case of \( D = 1 \) the summands corresponding to \( l_\text{V} = 0 \) have to be replaced
	according to the rule
	\begin{align}
		\label{eq D=1 rule MB}
		\frac{ (\varrho_0 E)^{\frac{l_\text{S} (D-1)}{2} -1} }{ \Gamma \bigl( \frac{l_\text{S} (D-1)}{2} \bigr) } \theta(E)
			\stackrel{D\rightarrow1}{\longrightarrow}
			\delta(\varrho_0 E)
	\end{align}
	and in~(\ref{eq def gamma}) the surface \( S_0 \) has to be taken as the number of bordering end-points, which would be
	two for a finite line and zero for the one-sphere-topology.
	The unconfined expression~(\ref{eq rho sym unconfined 2}) can easily be reobtained as the special case
	\( \gamma = 0 \) by recognizing that \( C_{l,l_\text{V}=l} = C_l \).

	The highest power of the energy in~(\ref{eq rho sym unconfined 2},\ref{eq rho sym confined})
	has the exponent \( (N D) \slash 2 - 1 \), which reminds of the Thomas-Fermi
	approximation of the effectively \( N \cdot D \)-dimensional billiard.
	This is not surprising, since \( l = N \) refers to a partition into unities \( N = 1 + \cdots + 1 \) associated solely
	with the identity permutation \(  P = \text{id}_{S_N} = (\,1\,)\,(\,2\,)\,\cdots\,(\,N\,) \).
	In the geometrical picture, this corresponds to the propagation of individual particles.
	None of them are clustered and in~(\ref{eq rho sym confined}) none of them are reflected on the boundary,
	since \( l_\text{S} = 0 \) for the highest power.
	The combinatorial factor~(\ref{eq comb factor}) and the coefficients~(\ref{eq univ coeff unconfined})
	and~(\ref{eq univ coeff confined}) compute to \( c(1,\ldots,1) = C_N = C_{N,N} = 1 \) and the corresponding term
	in~(\ref{eq rho sym unconfined}), (\ref{eq rho sym unconfined 2}) and~(\ref{eq rho sym confined})
	is the volume Weyl term of the fundamental domain in \( N \cdot D \)-dimensional space
	\begin{align}
		\label{eq rho volume term}
		\begin{split}
			\bar{\varrho}_\te{v}(E) &= \frac{1}{N!} \varrho_0^{\frac{N D}{2}} \frac{E^{\frac{N D}{2}-1}}{\Gamma\left( \frac{N D}{2} \right)} \theta(E) \\
			&= \frac{1}{N!} \left( \frac{m}{2 \pi \hbar^2} \right)^\frac{N D}{2} \frac{V_D^N }{\Gamma\left(\frac{N D}{2}\right)}
				E^{\frac{N D}{2}-1} \theta(E) \\
			&= \frac{1}{N!} \bar{\varrho}_\te{TF}(E) \, .
		\end{split}
	\end{align}
	The next section will show the importance of all corrections in~(\ref{eq rho sym unconfined}),
	(\ref{eq rho sym unconfined 2}) and~(\ref{eq rho sym confined})
	beyond the volume term~(\ref{eq rho volume term}).
\section{The Geometrical Emergence of Ground State Energies in \\Fermionic Systems}
\label{sec geom emergence}%
	As often in the context of semiclassics the case of a two-dimensional billiard is of special interest.
	On the one hand, this is because of possible technical applications. One can think of confined two-dimensional
	electron gases in semiconductor heterostructures or two-dimensional superconducting structures with bosonic
	description due to Cooper pairing for example.
	On the other hand, the existence of equally distributed energies in a \( 2 D \) single-particle billiard without
	boundary corrections is a valuable special feature.
	This is not only because of the exceptionally simple form that the density of states takes in these systems.
	A constant single-particle smooth part also opens the possibility to make connections to number theory.
	Namely approximations for average distributions of partitions of integers can be related.

	For positive arguments \( E \) the unconfined DOS~(\ref{eq rho sym unconfined 2}) in \( D = 2 \) is a polynomial of degree \( N-1 \)
	in the energy with coefficients that are just rational numbers.
	The coefficients can be summed up exactly for explicit values of \( N \) and \( l \) albeit with computation time
	increasing very strongly with \( N \) when using the form at hand~(\ref{eq rho sym unconfined}) or~(\ref{eq rho sym unconfined 2}).
	Note that the form of the coefficients used in~(\ref{eq rho sym unconfined}) in terms of ordered partitions
	corresponds to less computation time while the form~(\ref{eq rho sym unconfined 2}) seems to be a better
	starting point for simplifications or analytical calculations.

\begin{figure}
      \begin{center}
      \subfloat[2 particles]{\label{fig:plot 2 p}\includegraphics[width=0.328\textwidth]{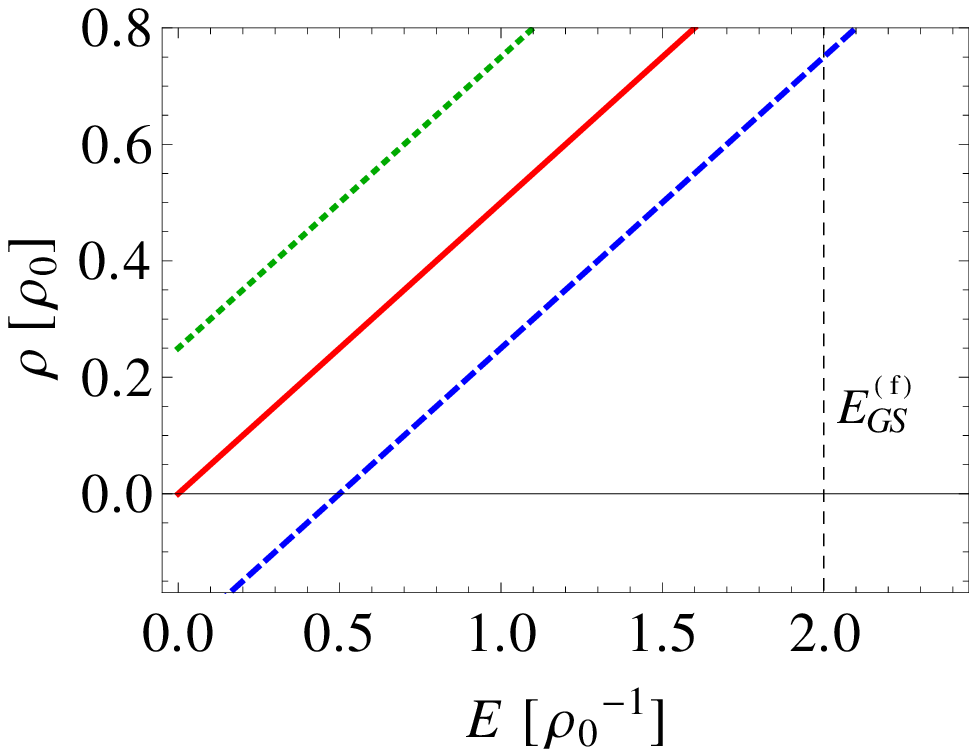}}\hfill
      \subfloat[3 particles]{\label{fig:plot 3 p}\includegraphics[width=0.328\textwidth]{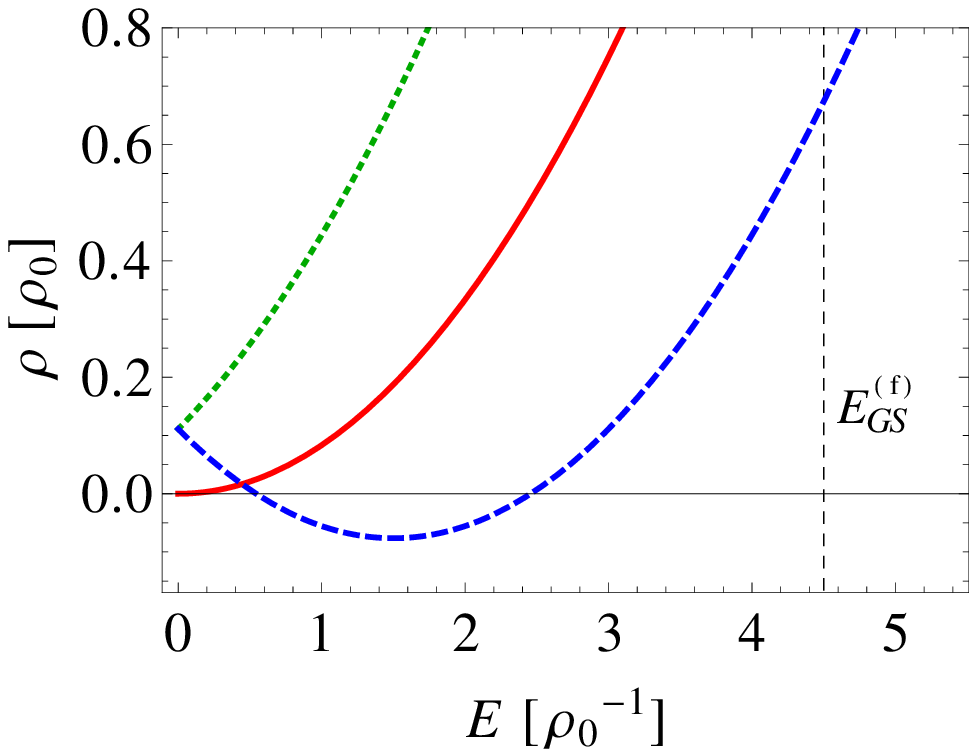}}\hfill
      \subfloat[4 particles]{\label{fig:plot 4 p}\includegraphics[width=0.328\textwidth]{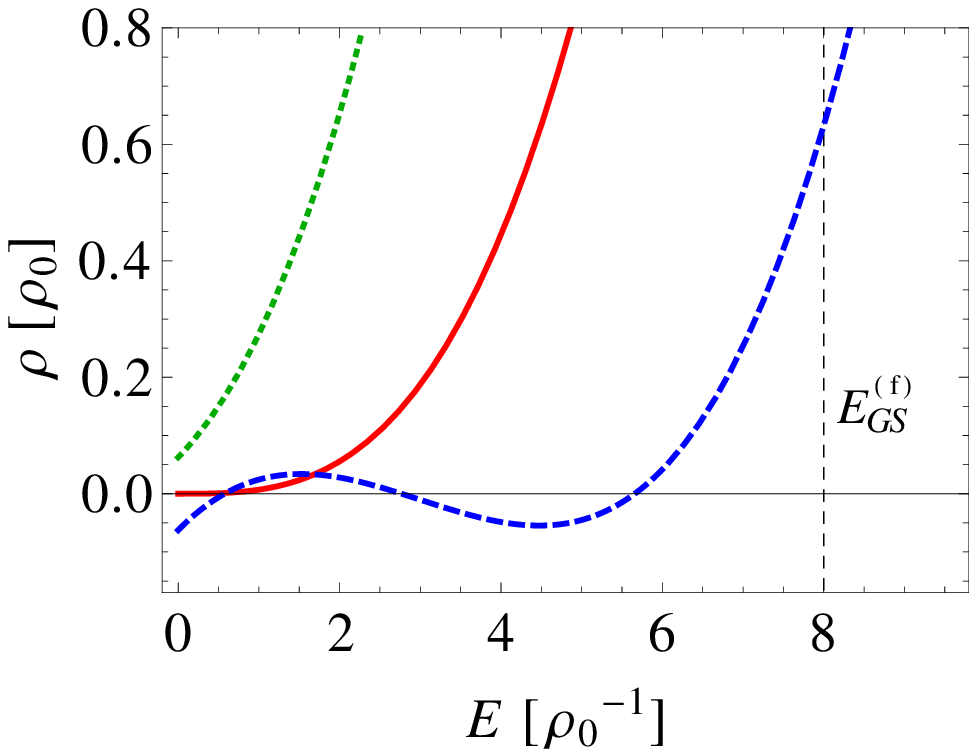}}
			\newline
      \subfloat[5 particles]{\label{fig:plot 5 p}\includegraphics[width=0.328\textwidth]{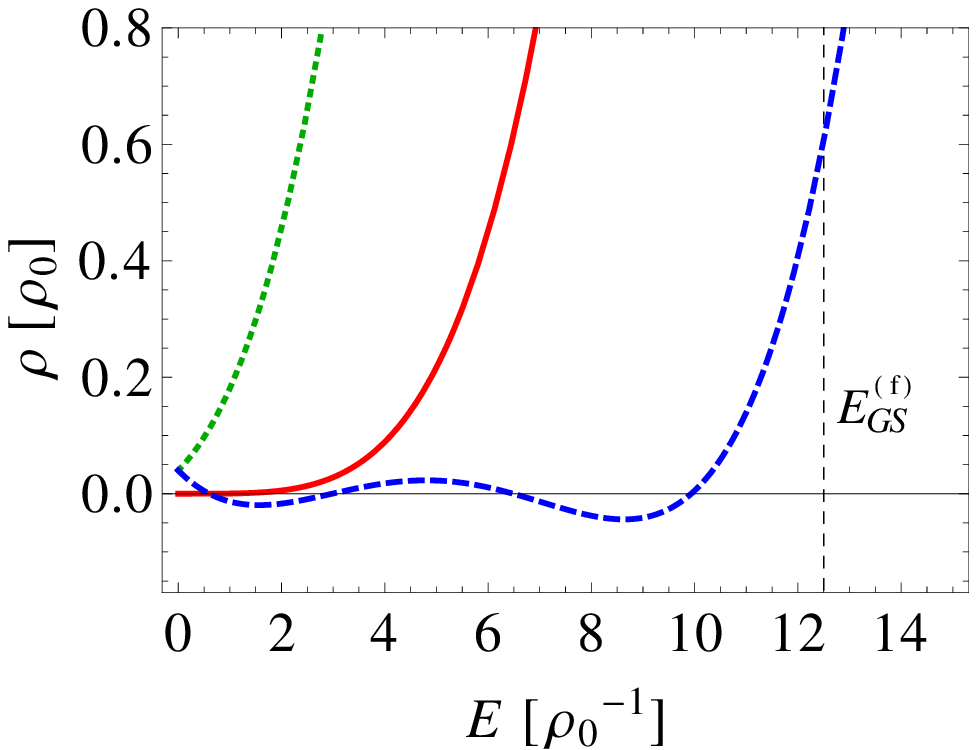}}\hfill
      \subfloat[7 particles]{\label{fig:plot 7 p}\includegraphics[width=0.328\textwidth]{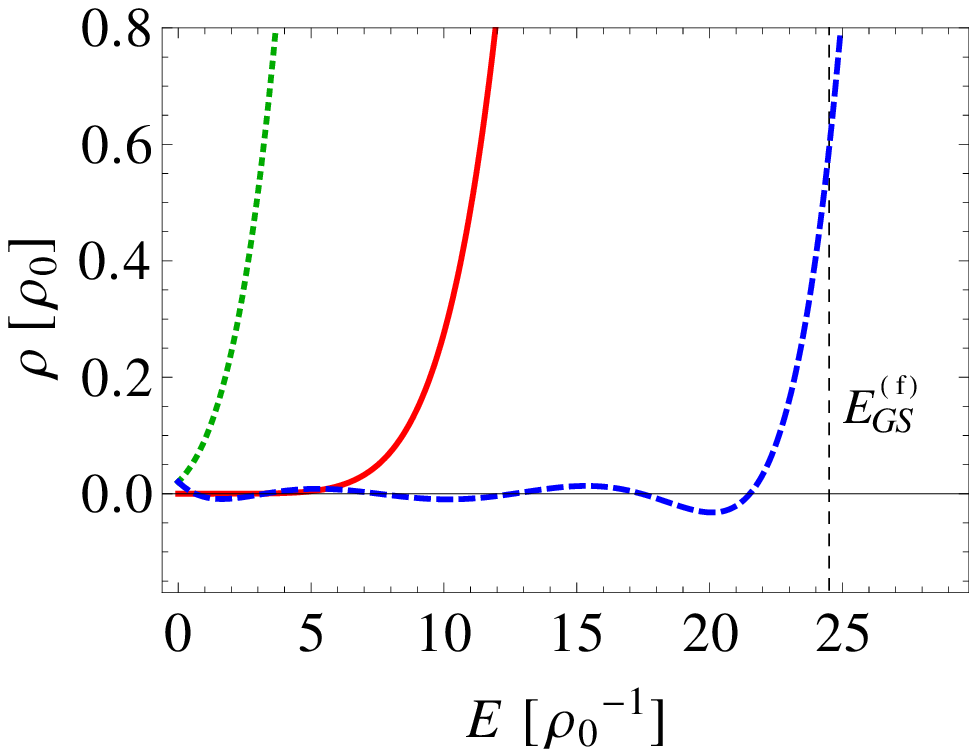}}\hfill
      \subfloat[20 particles]{\label{fig:plot 20 p}\includegraphics[width=0.328\textwidth]{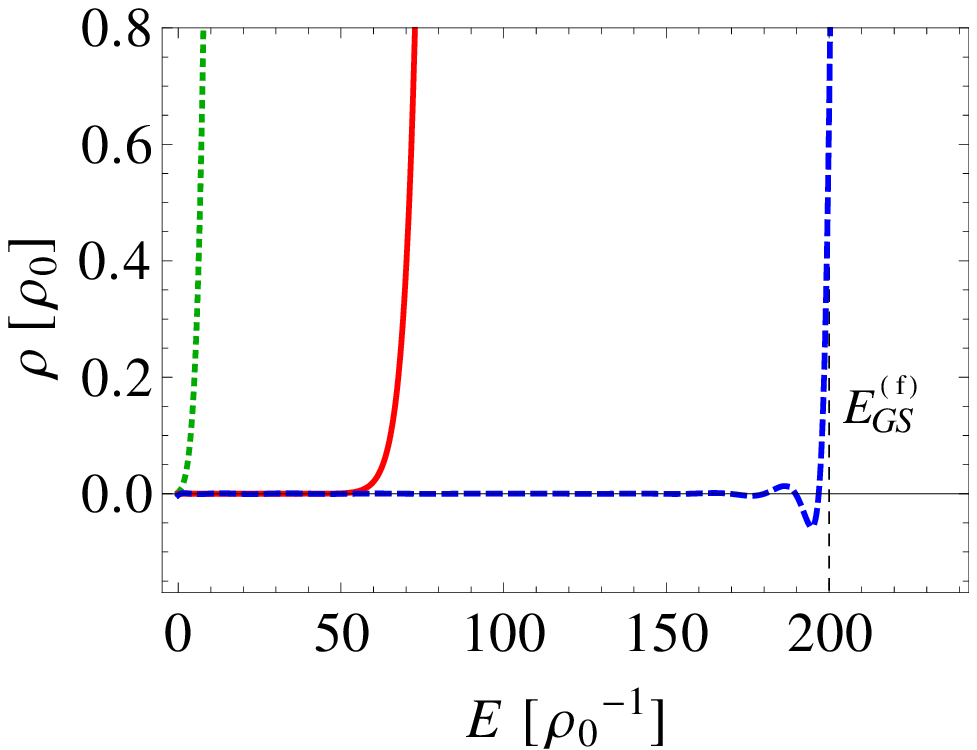}}
      \newline
      \caption{\label{fig:plot 2-20 p}Symmetry-projected DOS without boundary corrections in \( D=2 \)
				for bosons (green, dotted) and fermions (blue, dashed)
				in comparison to the naive volume term (red, solid) for several numbers of particles.
				The vertical dashed black line shows the particular expected many-body ground state energy
				\( E_\te{GS}^\te{(f)} \) for fermions~(\ref{eq E GS f}).
				Densities \( \varrho \) are measured in units of the constant single-particle density \( \varrho_0 \).
				Energies \( E \) are measured in units of its reciprocal \( \varrho_0^{-1} \)}
      \end{center}
    \end{figure}

	Figure~\ref{fig:plot 2 p} shows the case of two particles.
	The bosonic and fermionic cases are shown in comparison to the naive volume term~(\ref{eq rho volume term}).
    Already here, the symmetry corrections give qualitatively the right picture.
		With respect to the naive term, the fermionic density is shifted to higher energies, which is according to the
		expectation of the many-body ground state energy below which effectively no level should appear.
    Whereas the bosonic density is shifted to lower energies, which accords to the full counting of many-body levels corresponding
    to shared single-particle energies in contrast to the naive term, where these are counted with a factor of \( 1 \slash N! \),
		even if they can not be permuted in \( N! \) ways due to identity of some of the single-particle energies.

    Figures~\ref{fig:plot 3 p}--\ref{fig:plot 20 p} show the cases of two to twenty particles.
		In the fermionic case the lower powers in \( E \) in the polynomial produce oscillations around the axis \( \bar{\varrho} = 0 \).
		With increasing particle number these oscillations get smaller in amplitude and larger in number, approximating a zero-valued DOS.
		The DOS is effectively shifted to higher energies and an energy gap opens that coincides with the expected fermionic
		ground state energy \( E_\te{GS}^\te{(f)} \) calculated by counting single-particle levels by virtue of the smooth single-particle
		DOS \( \bar{\varrho}_\spp(E) \).
		It is defined by
		\begin{align}
			\label{eq E GS f}
			E_\te{GS}^\te{(f)} &:= \int_{-\infty}^{\bar{E}_\te{F}} \dd E' \, \bar{\varrho}_\spp(E') E' \, , \\
		\intertext{where the Fermi energy \( \bar{E}_\te{F} \) is defined through}
			\label{eq E Fermi}
			N &=: \int_{-\infty}^{\bar{E}_\te{F}} \dd E' \, \bar{\varrho}_\spp(E') \, ,
		\end{align}
		but instead of explicitly filling up single-particle energy
		levels by hand, this time the ground state energy occurs as a consequence out of exchange symmetry incorporated as a modification
    of the propagator. The corrections from cluster zone propagations are sufficient to automatically reproduce the expected ground
    state energy.
		When increasing \( N \), the symmetry projected DOS at \( E = E_\te{GS}^\te{(f)} \) is observed to keep moderate values
		\( \bar{\varrho}_-(E_\te{GS}^\te{(f)}) \approx \mathcal{O}(1) \) while the naive density at this energy grows exponentially with \( N \).
		In contrast to the fermionic density, the bosonic density does not have these oscillations. There, the polynomial
    in \( E \) has only positive coefficients and the density is effectively shifted to lower energies as expected intuitively.

\begin{SCfigure}
			\includegraphics[width=0.57\textwidth]{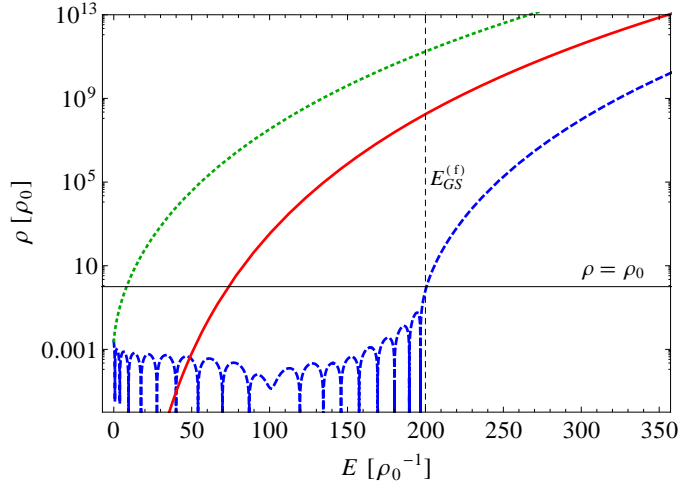}
			\caption{\label{fig:plot 20 p unconfined} Symmetry-projected DOS in \( D=2 \) for bosons (green, dotted) and
				fermions (blue, dashed) in comparison to the naive volume term (red, solid) for \( N=20 \) particles.
				Negative values of the fermionic density are plotted as logarithm of its absolute value.
				The expected many-body ground state energy \( E_\te{GS}^\te{(f)} \)
				for fermions~(\ref{eq E GS f}) is represented by the dashed vertical line.}
\end{SCfigure}%
	For higher particle numbers a single logarithmic plot suggests itself in order to show at the same time
	the oscillations that are becoming very small and the strong growth behaviour around and above the ground state energy.
	Figure~\ref{fig:plot 20 p unconfined} shows the smooth part of the density in the case of \( N = 20 \) particles.
	Again the fermionic energy gap accurately reproduces the ground state energy, indicated by crossing the axis of abscissa.
	Also in the bosonic case, the corresponding density \( \bar{\varrho}_+(E) \) apparently keeps moderate values at the
	expected bosonic ground state energy \( E_\te{GS}^\te{(b)} \) computed in analogue manner to \( E_\te{GS}^\te{(f)} \).

\begin{figure}[h]
      \begin{center}
				\subfloat[Absolute value of the many-body DOS for \(N = 13\) fermions.
					The blue solid curve shows the full expression~(\ref{eq rho sym unconfined}).
					The black curves correspond to leaving out the contributions \( l=1 \) (dashed), \(l=2\) (dash-dotted) and
					\(l=3\) (dotted).]
					{\label{fig:plot 13 missing}\includegraphics[width=0.53\textwidth]{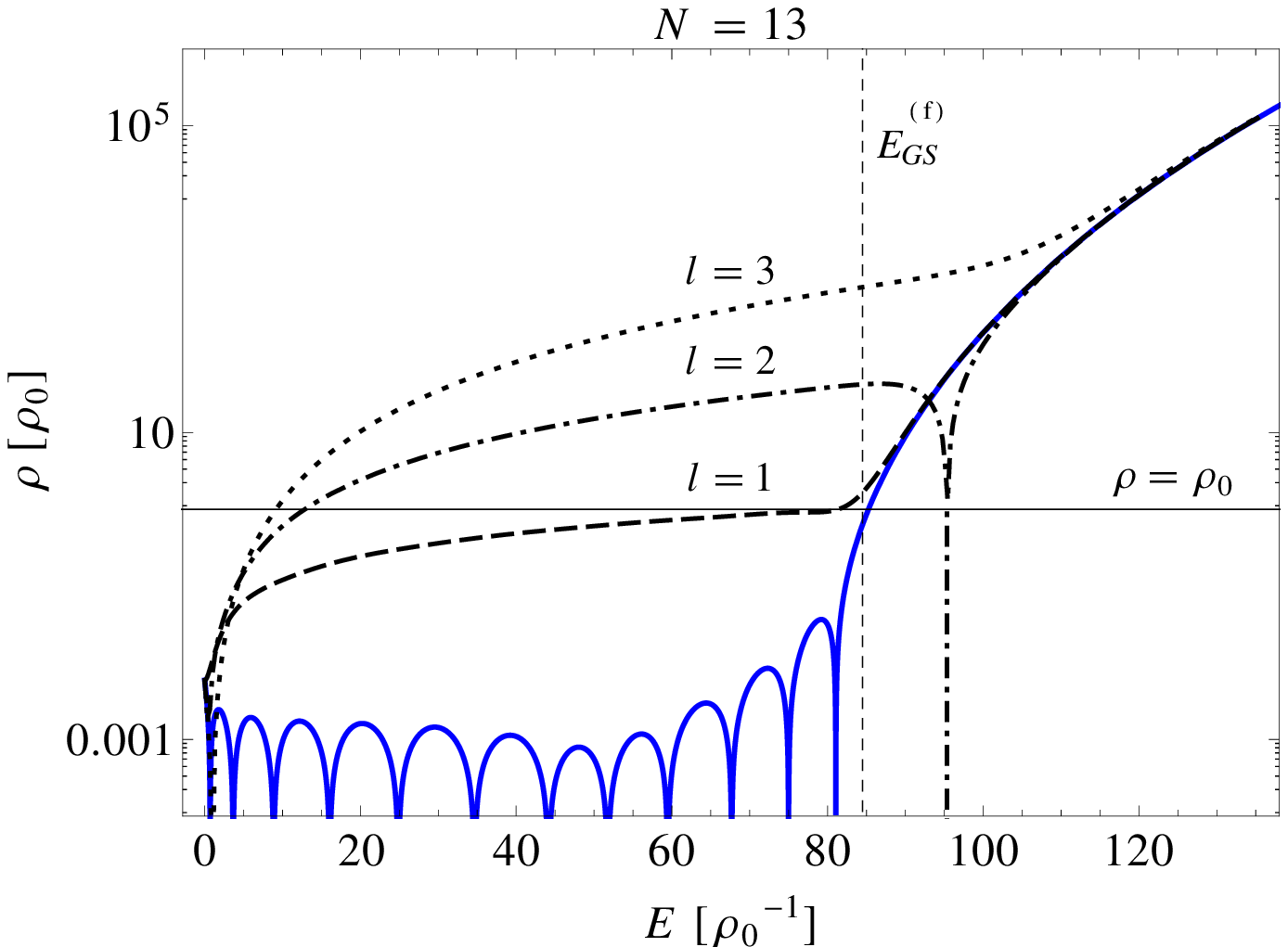}}\hfill
				\subfloat[Absolute value of multiple integrals
					\( \varrho^{(-n)}(E) = \int_0^E \dd e_n \cdots \int_0^{e_2} \dd e_1 \, \varrho_-(e_1)  \) of the many-body DOS
					for \(N = 17\) fermions. Oscillations around zero for energies below \( E_\te{GS}^\te{(f)}\) (vertical dashed line) remain.]
					{\label{fig:plot 17 integrals}\includegraphics[width=0.44\textwidth]{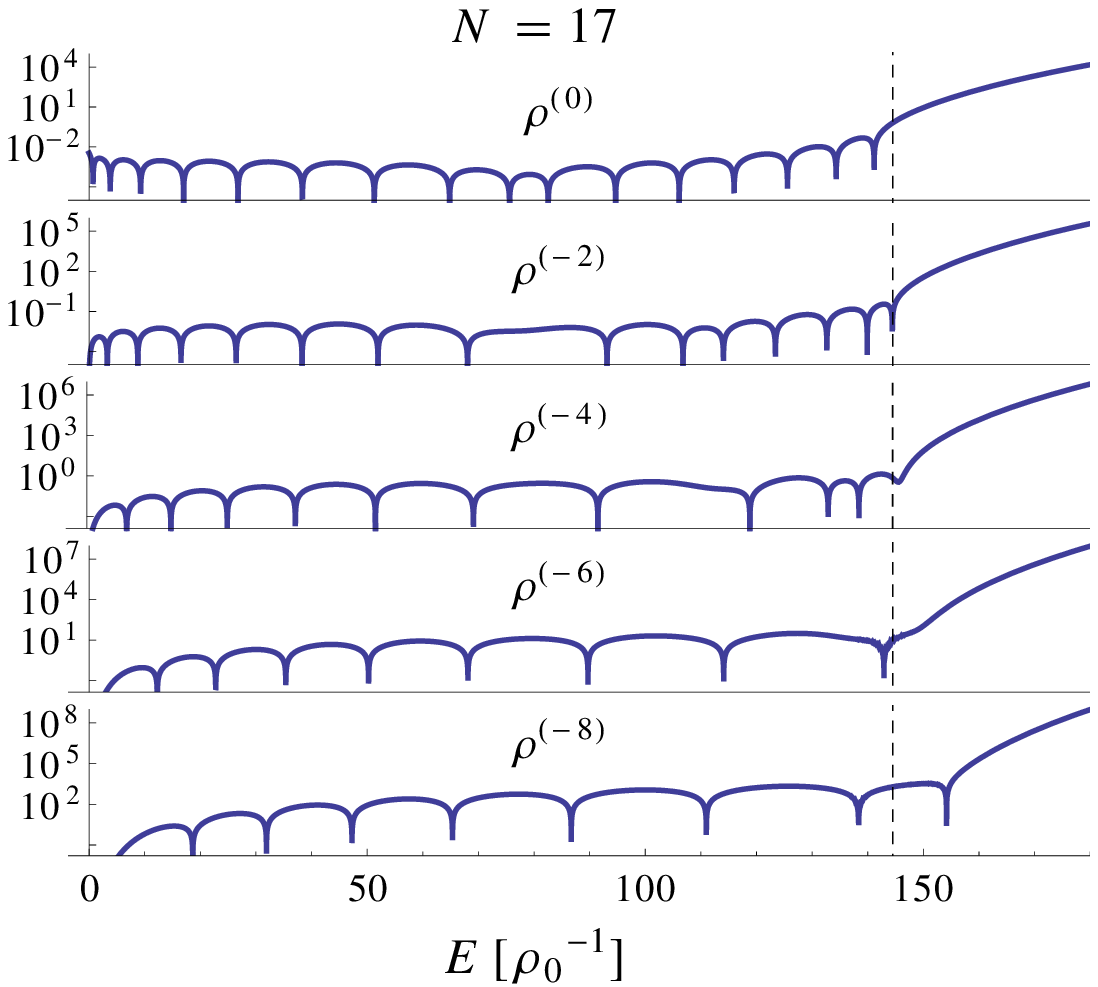}}\hfill
      \caption{Sensitivity (figure~\ref{fig:plot 13 missing}) respectively rigidity (figure~\ref{fig:plot 17 integrals}) of the oscillations of
				\( \bar{\varrho}_-(E) \).}
      \end{center}
    \end{figure}

	The small values of \( \bar{\varrho}_-(E) \) for \( E \lesssim E_\te{GS}^\te{(f)} \) result from large cancellations regarding
	the different terms in the sum~(\ref{eq rho sym unconfined 2}) with different values of \( l \). Therefore the behaviour of the DOS in this regime is very
	sensitive to numerical errors and all the corrections are needed to reproduce it correctly.
	In order to illustrate this fact figure~\ref{fig:plot 13 missing} shows the deviations one obtains
	when leaving out contributions.

\begin{figure}
		\psfrag{xaxis}[bc]{\small \(E [\varrho_0^{-1}] \)}
		\psfrag{yaxis}[tl][bl][1][180]{\small \( \mathcal{N}(E) \)}
		\psfrag{label}[bc]{\small \( N = 12 \)}
		\psfrag{EGS}[bl]{\scriptsize \( E_\te{GS}^\te{(f)} \)}
		\psfrag{EGSi}[bl]{\scriptsize \( E_\te{GS}^\te{(f)} \)}
		\psfrag{N12}[Bl]{\scriptsize \( \mathcal{N}= 1 \slash 2 \)}
		\psfrag{N12i}[Bl]{\scriptsize \( \mathcal{N}= 1 \slash 2 \)}
		\includegraphics[width=0.95\textwidth]{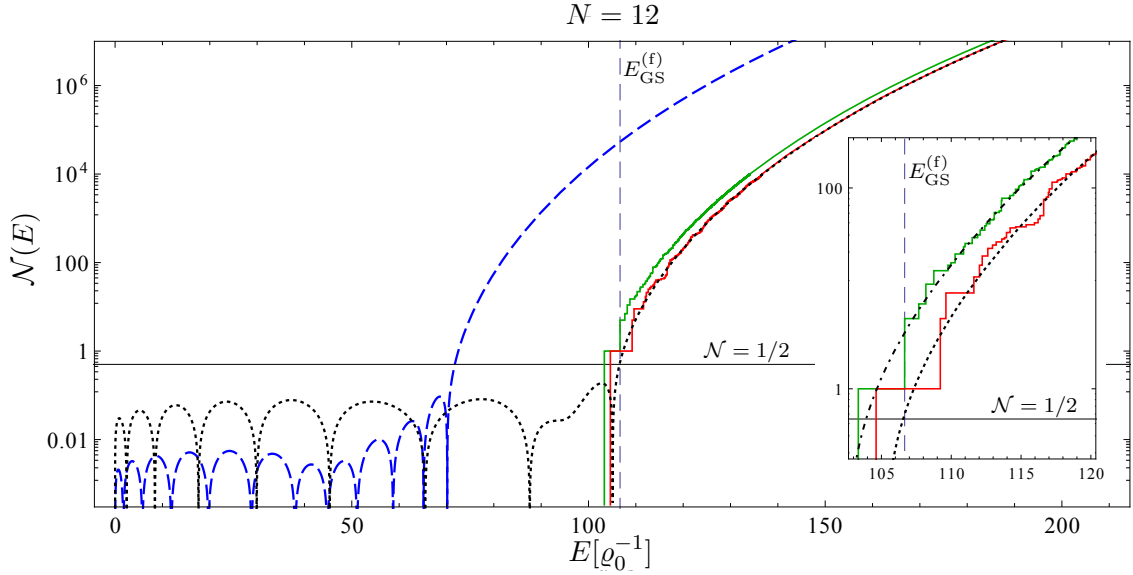}\hfill
		\caption{\label{fig:plot 12 p confined cylinder}Level counting function for 12 fermions.
			The blue dashed curve shows the smooth part without confinement corrections.
			The black dotted curve includes boundary corrections with a geometrical parameter of
			\( \gamma = - \sqrt{\pi} \slash 2 \), which corresponds to a circular billiard.
			The green staircase (left) shows the exact non-interacting many-body levels of the circular billiard.
			The red staircase (right) shows the exact levels of a cylindrical billiard with same geometrical parameter.
			The dash-dotted black curve in the inset shows the smooth part shifted to lower energies corresponding to the
			shift of \( E_\te{GS}^{(f)} \) due to curvature.}
	\end{figure}

	Another feature of the oscillations around zero up to \( E = E_\te{GS}^\te{(f)} \) is their rigidity with respect to integration
	in the sense that not only the DOS itself oscillates around zero but do also several integrals of it.
	Some integrals are shown in figure~\ref{fig:plot 17 integrals}.\\
	\\
	The effect of boundary corrections is shown in figure~\ref{fig:plot 12 p confined cylinder}.
	Displayed is the \( N=12 \) fermionic unconfined level counting function \( \mathcal{N}(E) = \int_0^E \dd E' \, \bar{\varrho}_-(E') \)
	in \( 2D \) and in addition to that the corresponding confined case~(\ref{eq rho sym confined}) for Dirichlet
	boundary conditions with a geometrical perimeter-to-area-ratio of \( \gamma = - \sqrt{\pi} \slash 2 \),
	which is the smallest possible parameter for singly connected billiards since it is the one of a circular billiard.
	The curve seems to be shifted further to higher energies, enlarging the energy gap.
	Again the expected ground state energy \( E_\te{GS}^{(f)} \), this time calculated using the single-particle Weyl expansion
	with perimeter correction, is reproduced very well.
	Already for this minimal value of \( \gamma \) the deviations from the unconfined case apparently are rather strong.
	Thus naturally the question arises whether the assumption of locally flat boundaries gives a sufficient description
	of an actual billiard or additional corrections also lead to rather strong deviations, making the treatment of curvature inevitable.
	Therefore the exact quantum mechanically solvable levels of a circular billiard have been arranged to non-interacting fermionic
	many-body levels shown by the green staircase function.
	The deviation seems again to be effectively a slight shift, smaller than the deviation of the confined from the unconfined case.
	Not displayed is \( E_\te{GS}^{(f)} \) including curvature, which in this case coincides with the exact circle levels, in the sense
	that it lies half way between the first two of them.
	The additional comparison with exactly computed levels of a billiard with the shape of a cylinder barrel
	with same \( \gamma \)-ratio as the circle serving as an example of a \( 2D \) system without curvature shows good agreement
	with the smooth part.
	Note that in this geometry the minimal value of \( \gamma \) could be underrun, which is not contradictory since it is not
	singly connected.
	The absence of curvature can be seen in the single-particle expansion given by Balian and Bloch~\cite{BalianBloch},
	where the corresponding correction is proportional to the Euler characteristic \( \chi \) of the billiard,
	which happens to be one for a disk and zero for a cylinder barrel.

	An analysis of the \( E_\te{GS}^\te{(f)} \) in \( 2D \) for the unconfined and confined cases with and without curvature
	indicate the relative importance of the corresponding contributions in the smooth DOS.
	The smooth ground state energy involving a Dirichlet type perimeter correction without curvature (\(\chi = 0\)) reads
	\begin{equation*}
		E_\te{GS}^\te{(f)}(N,\gamma,0) = E_\te{GS}^\te{(f)}(N,0,0) \, \left( \sqrt{1+a} + \sqrt{a} \right)^3
			\bigl( \sqrt{1+a} - \frac{1}{3} \sqrt{a} \bigr) \, , \qquad a = \frac{|\gamma|}{\sqrt{\pi N}} \, .
	\end{equation*}
	The inclusion of the curvature correction \( \chi \delta(E) \slash 6 \)~\cite{BalianBloch} in the single-particle
	DOS yields the ground state energy \( E_\te{GS}^\te{(f)}(N,\gamma,\chi) = E_\te{GS}^\te{(f)}(N-\chi \slash 6,\gamma,0) \).
	Comparing the correction due to the perimeter (\( \chi = 0 \)) with the further correction due to curvature
	\begin{equation*}
		\frac{ E_\te{GS}^\te{(f)}(N,\gamma,\chi) - E_\te{GS}^\te{(f)}(N,\gamma,0) }
			{ E_\te{GS}^\te{(f)}(N,\gamma,0) - E_\te{GS}^\te{(f)}(N,0,0) } = - \frac{ \sqrt{\pi} \chi }{ 8 | \gamma | }
			\sqrt{\frac{1}{N}} + \frac{\chi}{16 N} + \mathcal{O}\bigl( N^{-\frac{3}{2}} \bigr) \, ,
	\end{equation*}
	indicates that curvature contributions in general get strongly suppressed for large particle numbers.
	Based on this argument, the smooth part of the many-body DOS including curvature in \( D = 2 \) might be approximated by simply
	shifting the many-body DOS (including perimeter corrections) by
	\( \Delta E = E_\te{GS}^\te{(f)}(N,\gamma,\chi) - E_\te{GS}^\te{(f)}(N,\gamma,0) \).
	The corresponding function is plotted in the inset of figure~\ref{fig:plot 12 p confined cylinder}.
 \begin{figure}
      \begin{center}
      \psfrag{xaxis}[Bl]{\scriptsize \(E [\varrho_0^{-1}] \)}
			\psfrag{yaxis}[tl][bl][1][180]{\scriptsize \( \mathcal{N}(E) \)}
			{\psfrag{label}[bl]{\small \( N = 6 \)}
				\subfloat{\includegraphics[width=0.49\textwidth]{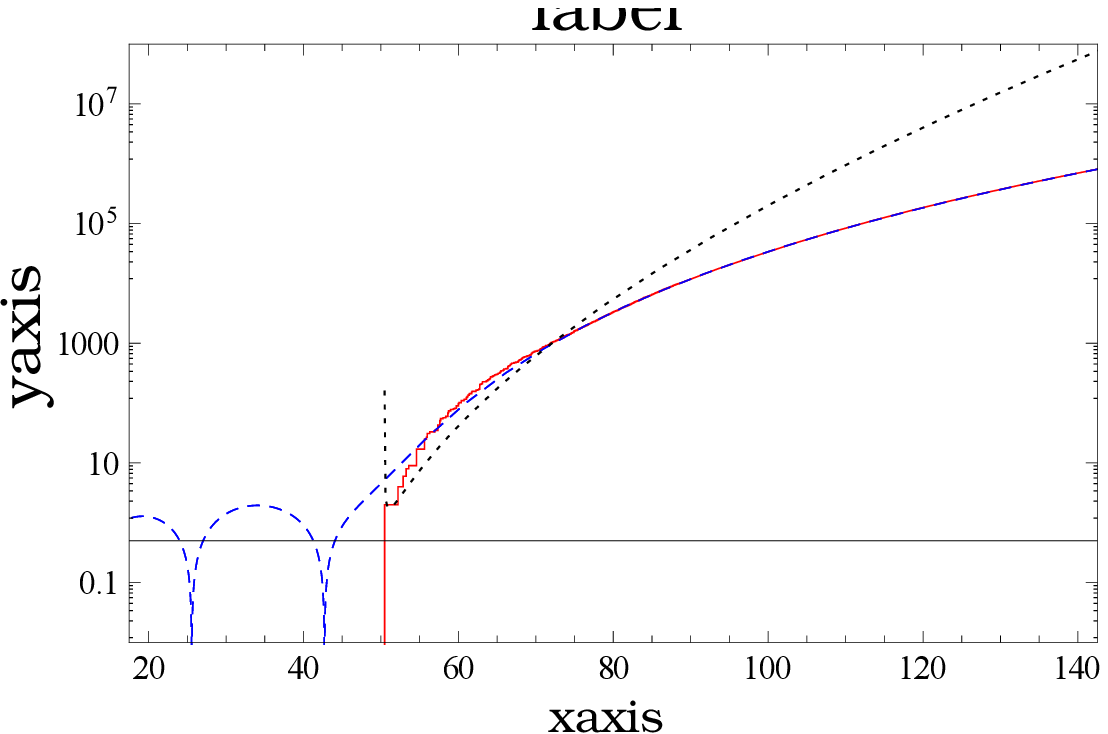}}}\hfill
      {\psfrag{label}[bl]{\small \( N = 9 \)}
				\subfloat{\includegraphics[width=0.49\textwidth]{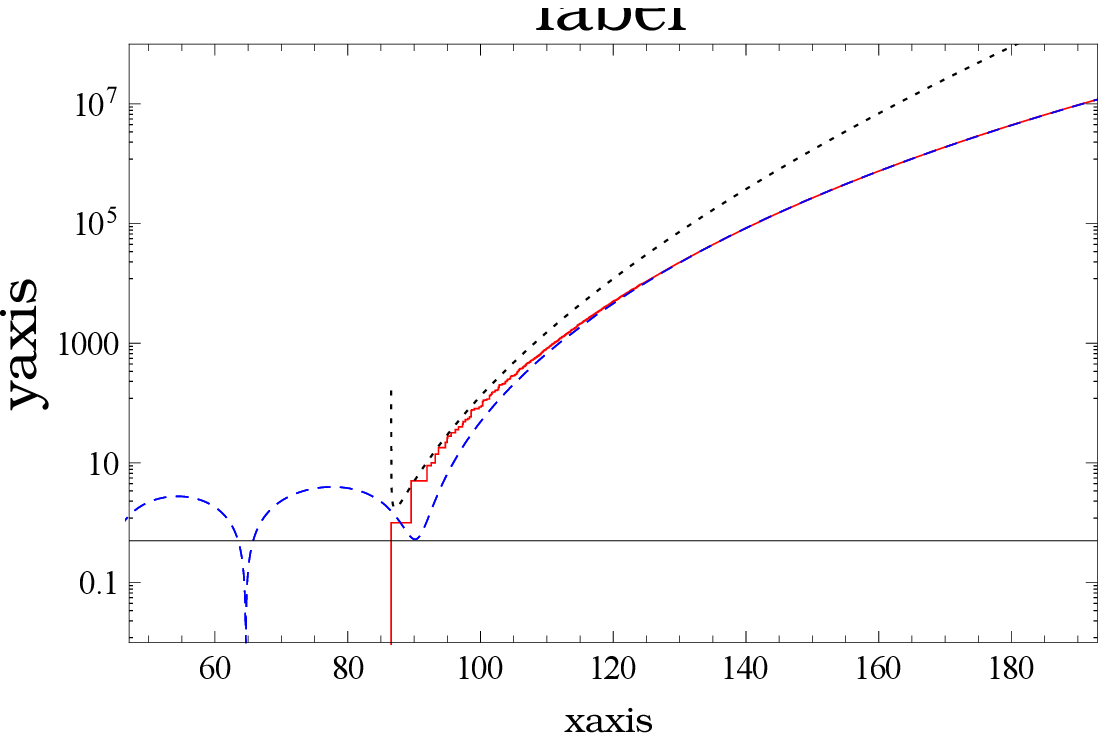}}}
      {\psfrag{label}[bl]{\small \( N = 12 \)}
				\subfloat{\includegraphics[width=0.49\textwidth]{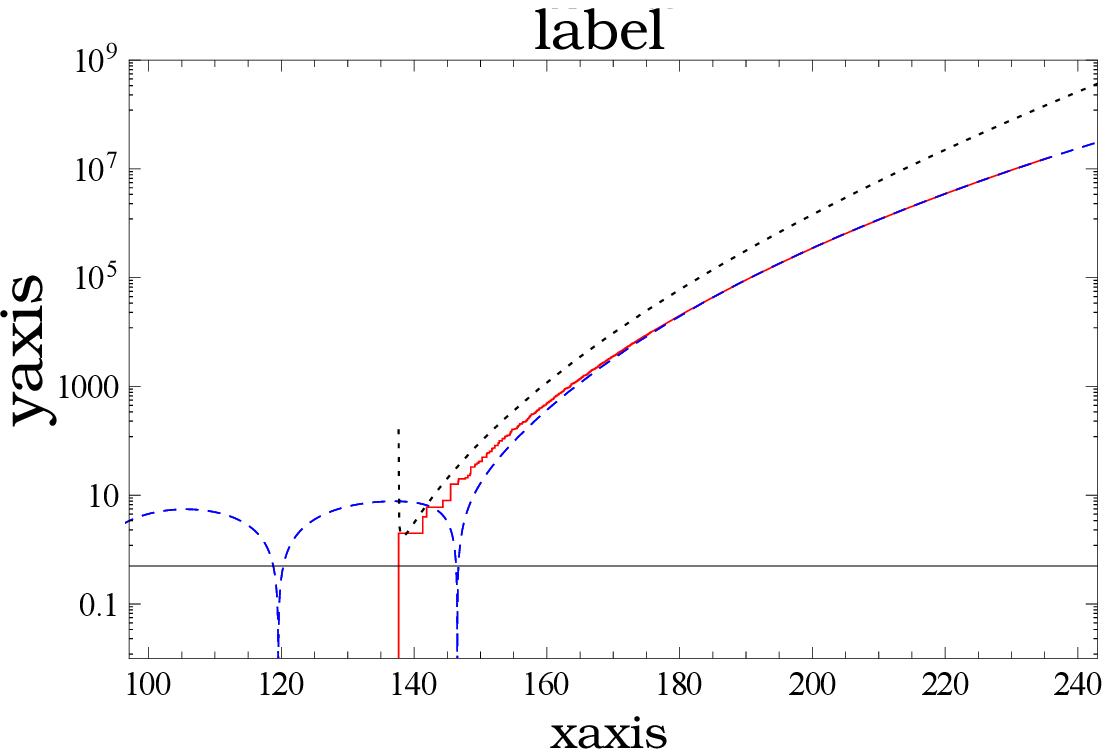}}}\hfill
      {\psfrag{label}[bl]{\small \( N = 16 \)}
				\subfloat{\includegraphics[width=0.49\textwidth]{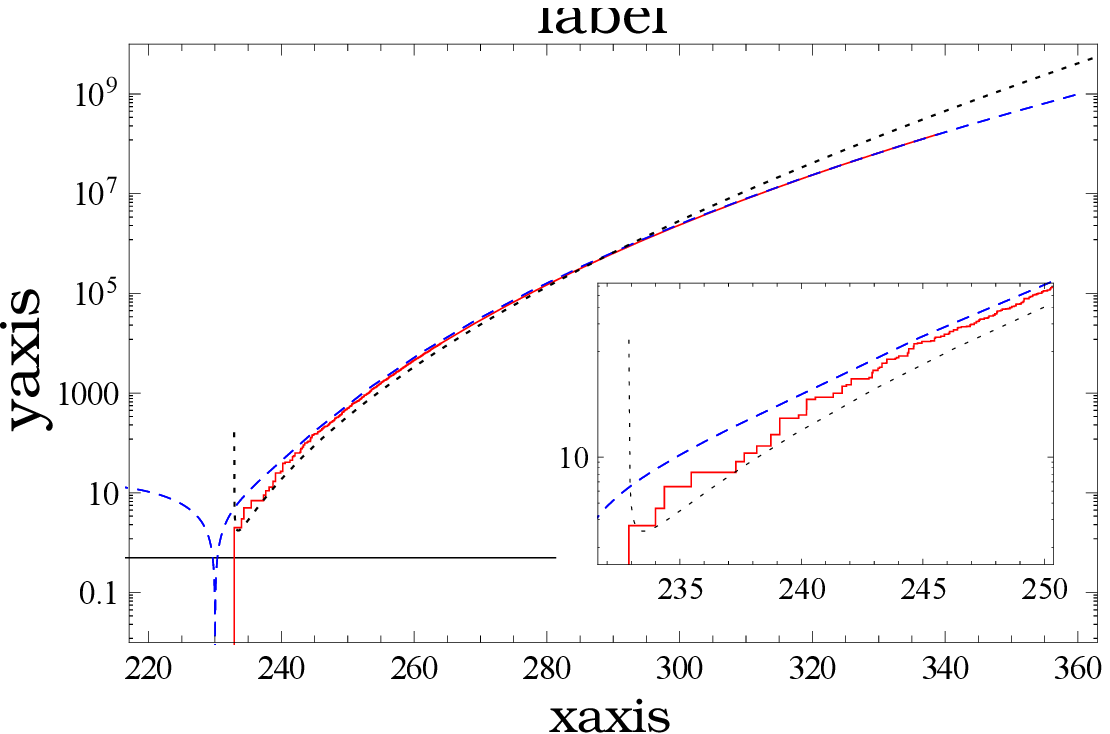}}}
      \caption{\label{fig:plot ring}Level counting function of a ring billiard with \(N=6,9,12,16\) fermions.
				The blue dashed curve shows the (absolute value of the) smooth part with confinement corrections
				(\ref{eq rho sym confined}) using a geometrical parameter of
				\( \gamma = - \sqrt{3 \pi} \slash 2 \).
				The red staircase shows exact non-interacting many-body levels of
				a ring billiard with the same geometrical parameter which corresponds to a ratio of radii of \( 2 \).
				The black dotted curve displays the Bethe approximation~(\ref{eq rho Bethe}) using the exact ground state energy as
				input.
				The horizontal black line corresponds to \( \mathcal{N} = 1 \slash 2 \).}
      \end{center}
    \end{figure}%
	It has to be emphasized that the ground state energies reproduced by the smooth densities are defined by smooth counting
	and do not necessarily have to accurately coincide with the exact many-body ground state energies of actual non-interacting quantum
	billiards.
	The exact fermionic ground state levels usually are subject to fluctuations around \( E_\te{GS}^\te{(f)} \) with respect to the
	particle number \( N \), which can be treated with semiclassical methods including corrections from periodic orbits~\cite{Centelles}.
	For rather regular systems possibly featuring degeneracies, these fluctuations in general can become strong and even the average
	\( \langle E_\te{GS}^\te{ex}(N) \rangle_N \) over some window in \( N \) can deviate from the smooth prediction
	\( E_\te{GS}^\te{(f)}(N) \)~\cite{Centelles}.
	A related feature of exact non-interacting many-body spectra are fluctuations of the many-body DOS at low excitation respectively temperature
	which can also be reproduced as semiclassical shell corrections to the Bethe estimate~(\ref{eq rho Bethe}) as shown by
	Leboeuf \etal~\cite{LeboeufFluc}.
	Therefore, the smooth part of the symmetry projected DOS~(\ref{eq rho sym unconfined 2},\ref{eq rho sym confined})
	presented in this work should not be regarded as an estimate for the exact ground state energy of a given system or the behaviour
	of the actual many-body DOS at energies close to it.
	Nevertheless, as fluctuations tend to wash out for higher excitations, the correct behaviour is very well reproduced for energies
	above a critical excitation (which depends on the specific geometry of the system and the particle number \( N \)).
	Note that the theoretical prediction of this more universal behaviour does not need any further system specific input than the
	volume and surface (and a possible non-zero valued Euler characteristic) of the billiard.

	Figure~\ref{fig:plot ring} shows both, deviations of the smooth prediction near the ground state as manifestation
	of such fluctuations and the increasingly good agreement for higher energies using the example of a ring billiard with several
	numbers of fermions.
	Like the cylindrical billiard, this system has an Euler characteristic of zero and therefore there are no signatures of curvature.
	Compared are the smooth level counting function \( \mathcal{N}(E) \) in the confined case and
	the exact many-body counting function of a ring billiard with a ratio of \( R \slash r = 2 \) of outer and inner radius,
	which corresponds to a geometrical parameter of \( \gamma = - \sqrt{3 \pi} \slash 2 \).
\section{Comparison with Known Results}
\label{sec comp w known results}%
 \subsection{Smoothing in Weidenm\"uller Convolution Formula}
		As Weidenmüller pointed out~\cite{Weidenmueller}, the symmetry projected part of the exact DOS of a non-interacting system
		can be constructed as a sum of convolutions of the single-particle DOS with modified energies.
		The DOS for a non-interacting bosonic (fermionic) system with exact single-particle DOS \( \varrho_\spp(E) = \sum_i \delta(E-\epsilon_i) \)
		reads
		\begin{align}
      \label{eq weidenmuller conv}
      \begin{split}
				\varrho_\pm(E) = &\frac{1}{N!} \sum_{l=1}^N (\pm 1)^{N-l} \sum_{\substack{N_1,\ldots,N_l =1\\ N_1 \leq \cdots \leq N_l}}^N
					\delta_{N, \, \sum N_\omega} \;\; c(N_1,\ldots,N_l) \Bigl( \prod_{\omega=1}^l \frac{1}{N_\omega} \Bigr) \times \\
				&\int \dd E_1 \ldots \dd E_l \; \delta \Bigl( E-\sum_{\omega=1}^l E_\omega \Bigr)
				\;  \left[ \prod_{\omega=1}^l \varrho_\spp\Bigl( \frac{E_\omega}{N_\omega} \Bigr) \right] .
      \end{split}
    \end{align}
		The derivation of~(\ref{eq weidenmuller conv}) is on the one hand based on the fact that the non-interacting
		many-body propagator decomposes into a product of single-particle propagators. On the other hand, this form requires
		the convolution property
		\begin{align}
			\label{eq conv property K}
			\int \dd^D q' \, K^{(\spp)}({\bf q}'', {\bf q}'; t_2) K^{(\spp)}({\bf q}', {\bf q}; t_1) = K^{(\spp)}({\bf q}'', {\bf q}; t_1 + t_2)
		\end{align}
		of the single-particle propagator, which is strictly fulfilled for the exact quantum propagator associated with the exact
		DOS. Inserting the single-particle spectrum directly into~(\ref{eq weidenmuller conv}), one can, in accordance to~\cite{Weidenmueller}
		show that~(\ref{eq weidenmuller conv}) is a formal way to express the construction of many-body levels out of filling up \( N \)
		single-particle levels under the condition that these should be counted correctly. So that no two filled fermionic single
		particle levels are the same and bosonic states where any of the filled single-particle levels are the same are not undercounted
		by the factor \( 1 \slash N! \) but instead fully addressed by counting each of the corresponding possible distributions of
		single-particle energies among the \( N \) particles with the factor of \( (\prod_\omega N_\omega!) \slash N! \).
		Here, the \( N_\omega \) denote the numbers of particles sharing a single-particle level in the questioned many-body state.

		The question of whether and, if yes, to what extent formula~(\ref{eq weidenmuller conv}) can be utilized in order to obtain or
		compare to the smooth part of the symmetry projected DOS can be answered in a twofold way.
		Naturally this question reduces to the question of how reasonable it is to simply use the smooth part of the single-particle DOS
		in~(\ref{eq weidenmuller conv}) instead of the exact one.

		First, one could regard the smooth part of a spectrum as a convolution with some smearing function.
		The question is then, if the smoothing of the single-particle DOS in~(\ref{eq weidenmuller conv}) yields the same result as if
		one would smooth the many-body DOS directly by a similar smoothing procedure.
		It turns out that this can be answered positively if the function used for smoothing obeys two conditions.
		Let \( \delta_\alpha(x) \) denote the smoothing function with a continuous parameter \( \alpha \) describing its sharpness
		and \( \int_{-\infty}^\infty \dd x \, \delta_\alpha(x) = 1 \). If this function fulfils
		\begin{align}
			\label{eq delta alpha 1}
			\delta_{c \alpha}(x) = \frac{1}{c} \delta_\alpha\left( \frac{x}{c} \right) \, , \qquad c > 0 \, , \\
			\label{eq delta alpha 2}
			\int_{-\infty}^\infty \dd y \, \delta_\alpha(x-y) \delta_\beta(y) = \delta_{\alpha + \beta}(x) \, ,
		\end{align}
		then the replacement of the single-particle DOS \( \varrho_\spp \) by its convolution with \( \delta_\alpha \) yields
		the same as convolving the many-body DOS \( \varrho_\pm \) with \( \delta_{N \alpha} \).
		If the function \( \delta_\alpha \) perceived as a distribution had a finite mean value and variance, then
		(\ref{eq delta alpha 1}) implies that \( \alpha \) is proportional to the standard deviation.
		But then this would contradict the second condition~(\ref{eq delta alpha 2}) as the variance in general is additive under
		convolution instead of the standard deviation. This shows that a proper smoothing function in this context should not have
		a finite standard deviation.
		Nevertheless, there is a distribution without standard deviation fulfilling the requirements at hand, namely the Cauchy distribution
		\begin{align}
			\delta_\alpha(x) = \frac{\alpha}{\pi ( \alpha^2 + x^2 )} \, .
		\end{align}
		\\
		The second way to address the question is to replace the exact single-particle DOS in~(\ref{eq weidenmuller conv}) by the smooth
		part \( \bar{\varrho}_\spp \) defined by the Weyl expansion up to a specified order.
		In the cases of restricting to the volume term and considering also the first boundary correction in \( \bar{\varrho}_\spp \)
		the corresponding integrals over single-particle energies are carried out successively yielding a solvable recursion relation.
		\ref{app calc conv unconfined} and \ref{app calc conv confined} show these calculations with the outcome that this procedure yields the same result for \( \bar{\varrho}_\pm(E) \)
		as the investigation of the propagation in cluster zones~(\ref{eq rho sym unconfined 2},~\ref{eq rho sym confined}).

		The reason why both procedures give the same result lies in the requirement~(\ref{eq conv property K}).
		For the exact quantum propagator the fulfilment of it is obvious.
		In the unconfined case the exact propagator is replaced by the free propagator, which also obeys the convolution property,
		but only if the intermediate coordinates \( {\bf q}' \) are integrated over full space \( \mathbb{R}^D \) and not only
		the interior \( \Omega \) of the billiard. Recalling the derivation of~(\ref{eq rho sym unconfined 2}), this
		corresponds exactly to the extension of the integration limits for all intermediate coordinates \( q_2,\ldots,q_n \)
		in~(\ref{eq trace free prop n A B}) allowing the evaluation of the integral as multidimensional Gaussian.
		Thus the assumption of infinite distance to the boundary in the analysis of cluster propagations in combination with the
		usage of the free propagator actually preserves the convolution property~(\ref{eq conv property K}).

		In the confined case the free propagator is modified by a reflection term
		\begin{align}
			K^{(\spp)}_0({\bf q}',{\bf q}; t) \rightarrow K^{(\spp)}_0({\bf q}',{\bf q}; t) \pm K^{(\spp)}_0( R{\bf q}', {\bf q}; t) \, ,
		\end{align}
		where \( R \) denotes the reflection with respect to the boundary regarded as locally flat and \( \pm \) refers to
		Neumann respectively Dirichlet conditions.
		This replacement yields the Weyl expansion including the surface correction.
		The analysis in \ref{app K conv prop confined} shows that the so constructed propagator in combination with the assumption of
		local flatness of the boundary also exhibits the convolution property~(\ref{eq conv property K}).
		This shows the equivalence of the computation \( \bar{\varrho}_\pm(E) \) in the confined case~(\ref{eq rho sym confined})
		via the convolution of single-particle Weyl expansions up to the first boundary correction and the computation
		by investigation of the propagation in cluster zones including reflections on the boundary assumed as locally flat.

		This equivalence gives rise to the question whether it is possible to
    relate corrections from propagation in cluster zones to the correction of delta peaks that is inherent to the exact convolution
    formula~(\ref{eq weidenmuller conv}) of Weidenm\"uller.
		And indeed one can relate each cluster zone correction to the correction of delta peaks for total energies.
		The corrected total energy is the energy that is a partition of single-particle energies just
    the way the cluster zone corresponds to a partition of all particles into clusters.
		The cluster zone correction is associated to the correction of the total energy that is built of single-particle energies,
		where all particles in a cluster share the same energy.
		But of course it is doing it in a smooth manner, meaning it produces a smooth correction to the DOS corresponding to a
		smooth version of the density constructed out of such many-body energies with correct prefactor.
\subsection{Connection to Bethe's Estimate in Fermionic Systems}
		This section is dedicated to the connection between the expressions presented in this work and the celebrated Bethe approximation to
		the many-body DOS.
		As the latter results as a saddle point approximation of an integral representation of the DOS obtained
		by general quantum statistical methods for non-interacting fermionic systems,
		the link between both is the equivalence of the smooth part of the DOS~(\ref{eq rho sym unconfined 2},\ref{eq rho sym confined}) and the standard statistical formulation in terms of the single-particle DOS.
		This connection will be drawn by the usage of the technique of generating functions as the easiest to handle statistical
		object associated with the DOS is the grand canonical partition function \( \mathcal{Z}^\mathrm{GC}(z,\beta) \) which itself
		is the generating function of the canonical partition functions for \( N \) particles
		\begin{align}
			\mathcal{Z}^\mathrm{C}(N,\beta) = \frac{1}{N!} \frac{\dd^N}{\dd z^N} \mathcal{Z}^\mathrm{GC}(z,\beta) \Bigr\rvert_{z=0} \, .
		\end{align}
		The essential step in order to find the generating function for the DOS~(\ref{eq rho sym confined})
		\begin{align}
			G_\pm(z,E) &= \sum_{N=0}^\infty \bar{\varrho}_\pm(N,E) z^N, \\
			\Leftrightarrow \bar{\varrho}_\pm(N,E) &= \frac{1}{N!} \frac{\dd^N}{\dd z^N} G(z,E) \Bigr\rvert_{z=0} \, ,
		\end{align}
		is to represent the difficult to handle combinatorial numbers \( C_l, C_{l,l_\text{V}} \)
		(\ref{eq rho sym unconfined}),~(\ref{eq rho sym confined}) in terms of their generating functions.
		In order to ease notation, the abbreviations \( \mu = \frac{D}{2}+1 \) and \( \nu = \frac{D-1}{2}+1 \) are used.
		\begin{align}
			C_l^{(N)} = \frac{1}{N!} \frac{\dd^N}{\dd z^N}
				\left. \biggl( \sum_{n=0}^\infty \frac{z^n}{n^\mu} \biggr)^l
				\right|_{z=0}
				= \frac{1}{N!} \frac{\dd^N}{\dd z^N}
				\left. \left( \mathrm{Li}_\mu(z) \right)^l
				\right|_{z=0}
		\end{align}
		and similarly
		\begin{align}
			\label{eq gen func C 2}
			C_{l,l_\text{V}}^{(N)} &= \frac{1}{N!} \frac{\dd^N}{\dd z^N}
				\left. \left[
					\biggl( \sum_{n=0}^\infty \frac{z^n}{n^\mu} \biggr)^{l_\text{V}}
					\biggl( \sum_{n=0}^\infty \frac{z^n}{n^\nu} \biggr)^{l-l_\text{V}}
				\right] \right|_{z=0} \nonumber \\
			&=\frac{1}{N!} \frac{\dd^N}{\dd z^N}
				\left[
					\left( \mathrm{Li}_\mu(z) \right)^{l_\text{V}}
					\left( \mathrm{Li}_\nu(z) \right)^{l-l_\text{V}}
				\right] \Biggr\rvert_{z=0} \, .
		\end{align}
		These relations can be seen by factoring out the powers of the sums in parentheses and recognizing that the \(N\)-th
		power of \( z \) in the product must be built of \( l \) factors \( z^{n_i}, i=1,\ldots,l \) with \( \sum_i n_i = N \).
		Thus each such combination corresponds to a particular partition of \( N \) into \( l \) parts \( n_1 + \cdots + n_l \).
		The correct coefficients are then created as products of the parts \( n_i^{\mu \slash \nu} \) appearing as denominators
		in the sums over \( n \).
		The generating functions can also be built constructively which will be shown exemplarily for the unconfined case.
		The confined case can be treated similarly.
		For this calculation the form of \( C_l^{(N)} \) expressed as sum over distinct partitions instead of ordered tuples
		will be used.
		Note that \( C_l^{(N)} \) should be understood to be zero for all \( l > N \).
		The starting point is to calculate constructively the generating function of \( C_l^{(N)} \slash l! \) which is
		\begin{align}
			g_l(z) := \frac{1}{l!} \sum_{N=0}^\infty C_l^{(N)} z^N &= \sum_{N=0}^\infty \,
				\sum_{{\substack{N_1 \leq \ldots \leq N_l =1\\ \sum N_\omega = N}}}^N
				\biggl( \prod_{n=1}^\infty \frac{1}{m_n!} \biggr)
				\biggl( \prod_{\omega=1}^l \frac{1}{N_\omega} \biggr)^\mu z^N \nonumber \\
			\label{eq gen func C constr}
			&= \sum_{\substack{ {\bf m} \in \mathbb{N}_0^\mathbb{N} \\ \sum m_n = l}}
				\prod_{n=1}^\infty \frac{z^{n \, m_n}}{m_n! \, n^{\mu \, m_n}} \, ,
		\end{align}
		where the sum over all possible distinct partitions is expressed as sum over all multiplicities \( m_n \) with which the part
		sizes \( n=1,2,\ldots \) appear.
		Given a fixed value of \( l \) corresponds to the restriction \( \sum m_n = l \) in the sum.
		This restriction is dropped by constructing a further generating function of~(\ref{eq gen func C constr}), this time with respect to \( l \).
		\begin{align}
			\sum_{l=1}^\infty g_l(z) y^l = \prod_{n=1}^\infty \left[ \sum_{m=0}^\infty
				\frac{1}{m!} \left( \frac{z^n y}{n^\mu} \right)^m \right]
				= \prod_{n=1}^\infty \exp \left( \frac{z^n y}{n^\mu} \right)
				= \exp \left[ y \mathrm{Li}_\mu (z) \right] \, .
		\end{align}
		Making use of this secondary generating function shows that
		\begin{align}
			g_l(z) = \frac{1}{l!} \frac{\dd^l}{\dd y^l} \exp \left[ y \mathrm{Li}_\mu (z) \right] \Bigr\rvert_{y=0}
				= \frac{1}{l!} \left[ \mathrm{Li}_\mu (z) \right]^l \, .
		\end{align}
		Note that because of \( \mathrm{Li}_\mu(z) = \mathcal{O}(z) \) as \( z \rightarrow 0 \), indeed the \( C_l \) vanish
		for \( l>N \).\\
		\\
		From here, the confined case will explicitly be treated, including the unconfined case as \( \gamma = 0 \).
		Using~(\ref{eq gen func C 2}) in~(\ref{eq rho sym confined}) gives
		\begin{align}
			\label{eq gen func}
			G_\pm(z,E) = \varrho_0 \sum_{l=0}^\infty (\pm 1)^l \sum_{\substack{l_\text{V}, l_\text{S}=0 \\ \sum l_i = l}}^l
				\frac{ [\mathrm{Li}_\mu(\pm z)]^{l_\text{V}} [\mathrm{Li}_\nu(\pm z)]^{l_\text{S}} }
				{l_\text{V}! \; l_\text{S}! }
				\gamma^{l_\text{S}} \frac{ (\varrho_0 E)^{\lambda - 1}}
				{\Gamma\left( \lambda \right)} \theta(E)
		\end{align}
		with the abbreviated exponent \( \lambda = l_\text{V} D \slash 2 + l_\text{S} (D-1) \slash 2 \)
		and \( \mu = D \slash 2 + 1 \), \mbox{\( \nu = (D-1) \slash 2 + 1 \)} as before.
		The upper limit of the sum over \( l \) has been raised from \( N \) to infinity which is no problem
		since the \( N \)-th derivative at \( z=0 \) is not affected by terms with \( l > N \).
		Applying Laplace transformation switches from energy domain to the domain of inverse temperature
		and allows for exact summation of \( l_\text{V} \) and \( l_\text{S} \).
		\begin{align}
			\label{eq gen func inv L}
			G_\pm(z,E) &= \mathcal{L}^{-1}_\beta \bigg[ \sum_{l_\text{V}, l_\text{S} = 0}^\infty
				\frac{ [\pm \mathrm{Li}_\mu(\pm z)]^{l_\text{V}} [\pm \mathrm{Li}_\nu(\pm z)]^{l_\text{S}}}
				{l_\text{V}! \; l_\text{S}! } \gamma^{l_\text{S}} \Bigl( \frac{\varrho_0}{\beta} \Bigr)^\lambda \biggr](E) \nonumber \\
			&= \mathcal{L}^{-1}_\beta \left[ \exp \biggl( \pm \Bigl( \frac{\varrho_0}{\beta} \Bigr)^\frac{D}{2} \mathrm{Li}_\mu(\pm z)
				\pm \gamma \Bigl( \frac{\varrho_0}{\beta} \Bigr)^\frac{D-1}{2} \mathrm{Li}_\nu(\pm z) \biggr) \right](E) \, .
		\end{align}
		In the case of \( D = 1 \), the replacement rule~(\ref{eq D=1 rule MB}) is already accounted for in~(\ref{eq gen func inv L}).
		Note that the bilateral definition of Laplace transform is used, which simply yields the Heaviside step function \( \theta(E) \) after
		taking the inverse Laplace transform.

		The result~(\ref{eq gen func inv L}) is equivalent to the grand canonical potential one obtains by standard statistical
		methods.
		In general, the grand canonical potential of a non-interacting quantum system of identical particles can be expressed
		in terms of the single-particle DOS as
		\begin{align}
			\label{eq lnZGC}
			\ln \mathcal{Z}^\mathrm{GC}_\pm(z,\beta) = \mp \int \dd E \, \varrho_\spp(E) \ln(1 \mp z \mathrm{e}^{-\beta E}) \, ,
		\end{align}
		where the upper sign corresponds to bosonic symmetry (\( z \) has to be small enough in modulus to keep the argument of the
		logarithm inside the unit disk around \(1\)) and the lower one corresponds to fermionic statistics.
		If the single-particle DOS is taken as its smooth part built as a sum of smooth functions of the form
		\begin{align}
			\varrho_i(E) = \varrho_0 \gamma_i \frac{ (\varrho_0 E)^{\nu_i - 1} }{ \Gamma(\nu_i) } \theta(E) \, ,
		\end{align}
		each of the summands yields an additive contribution
		\begin{align}
			\label{eq lnZGC explicit}
			\ln \mathcal{Z}^\mathrm{GC}_{\pm, i}(z,\beta) &= \pm \gamma_i \varrho_0^{\nu_i}
				\sum_{k=1}^\infty \frac{(\pm z)^k}{k} \, \frac{1}{\Gamma(\nu_i)} \int_0^\infty \dd E \, E^{\nu_i-1} \mathrm{e}^{-k \beta E} \nonumber \\
				&= \pm \gamma_i \Bigl( \frac{\varrho_0}{\beta} \Bigr)^{\nu_i} \sum_{k=1}^\infty \frac{(\pm z)^k}{k^{\nu_i+1}} \nonumber \\
				&= \pm \gamma_i \Bigl( \frac{\varrho_0}{\beta} \Bigr)^{\nu_i} \mathrm{Li}_{\nu_i +1} (\pm z) \, ,
		\end{align}
		which shows the equivalence to~(\ref{eq gen func inv L}) due to \( \mathcal{Z}^\mathrm{GC} = \mathcal{L}[ G ] \)
		when taking the appropriate single-particle DOS with \( \gamma_1 = 1, \nu_1 = D \slash 2 \) in the first term and
		\( \gamma_2 = \gamma, \nu_2 = (D-1) \slash 2 \) in the second one.
		The derivation of~(\ref{eq lnZGC}) involves the replacement
		\begin{align}
			\sum_i f(E_i) \rightarrow \int \dd E \, \varrho_\spp(E) f(E)
		\end{align}
		of the sum over single-particle eigenenergies, which are assumed to be discrete in the first place.
		Simply taking the smooth part of the single-particle DOS instead is therefore closely related to taking the smooth part
		in the convolution formula~(\ref{eq weidenmuller conv}) by Weidenmüller, which we showed to be equivalent to the actual
		smooth part of the many-body density obtained with the geometrical approach.

		For practical reasons it should be mentioned that in the unconfined case \( \gamma = 0 \) in \( D = 2 \), the inverse Laplace transform in
		(\ref{eq gen func inv L}) can be done explicitly to get
		\begin{align}
			G_\pm^{(D=2,\gamma=0)}(z,E) = \varrho_0 \sqrt{\pm \frac{\mathrm{Li}_2(\pm z)}{\varrho_0 E}}
				\; \mathrm{I}_1 \! \left( 2 \sqrt{ \pm \mathrm{Li}_2(\pm z) \varrho_0 E } \right) \, ,
		\end{align}
		which could also be seen directly from~(\ref{eq gen func}) after recognizing the power series of the modified Bessel function
		\( \mathrm{I}_1(x) \) with the two factorials \( l_\text{V}! \) and \( \Gamma(\lambda) = (l_\text{V}-1)! \) in the denominator
		(\(l_\text{S}=0\)).\\
		\\
		The Bethe approximation~\cite{Bethe} of the non-interacting fermionic many-body DOS is based on the general statistical relation
		(\ref{eq lnZGC}) and reads
		\begin{align}
			\label{eq rho Bethe}
			\bar{\varrho}_\mathrm{B}(Q) = \frac{1}{\sqrt{48} \, Q} \exp\left( \sqrt{\frac{2 \pi^2}{3} \bar{\varrho} \, Q} \right) \, ,
		\end{align}
		where \( Q = E - E_\mathrm{GS} > 0 \) is the excitation energy above the many-body ground state energy~(\ref{eq E GS f}) and
		\( \bar{\varrho} = \bar{\varrho}_\spp(E_\mathrm{F}) \) is the single-particle DOS at the Fermi energy~(\ref{eq E Fermi}).
		It results from a low temperature expansion and a simultaneous complex
		saddle point approximation in two integrals.
		The first one being the Bromwich integral achieving the inverse Laplace transform of \( \mathcal{Z}^\mathrm{GC} \) and the
		second one being a closed complex contour integral representing its \( N \)-th derivative with respect to \( z \).
		The low temperature expansion limits the validity of~(\ref{eq rho Bethe}) to low excitation energies \( Q \lesssim E_\mathrm{F} \)
		(range increases with \( N \))
		whereas the defect produced by the saddle point approximations becomes noticeable when approaching the ground state energy
		\( Q \lesssim \bar{\varrho}^{-1} \) and culminates in the divergence at \( Q = 0 \).\\
		Despite the fact that~(\ref{eq rho Bethe}) is valid in general and is not depending on the specific form of the single-particle density,
		it shall be sketched here how to obtain it explicitly for the unconfined case using expression~(\ref{eq gen func inv L}) respectively
		(\ref{eq lnZGC explicit}), since the explicit expressions can be of use investigating possible extensions to~(\ref{eq rho Bethe}) that
		are not general. Expressing the inverse Laplace transform and derivation with respect to \( z \) as integrals, the smooth part of the
		fermionic many-body DOS for \( N \) particles in a \( D \)-dimensional billiard (\( \nu = D \slash 2 \)) without confinement
		corrections reads
		\begin{align}
			\label{eq rho as int}
			\bar{\varrho}_-(N,E) = \left(\frac{1}{2 \pi i}\right)^2 \int_{\mathrlap{\Gamma_\beta}} \; \dd \beta
				\int_{\mathrlap{\Gamma_\alpha}} \; \dd \alpha \;
				\exp \left[  \beta E - \alpha N - \beta^{-\nu} \mathrm{Li}_{\nu + 1}( - \mathrm{e}^\alpha ) \right] \, ,
		\end{align}
		where \( \mathrm{e}^\alpha := z \) and energies are measured in units of \( \varrho_0 \).
		The contours are \( \Gamma_\beta: (\epsilon -i \infty, \epsilon + i \infty) \) with \( \epsilon \rightarrow 0^+ \)
		because of the essential singularity at \( \beta = 0 \) and
		\( \Gamma_\alpha: (a - i \pi, a + i \pi) \) with real \( a \leq 0 \), corresponding to a closed counterclockwise circular
		contour for \( z \) inside the unit circle, chosen because of the branch cut \( z \in (-\infty,-1] \) of \( \mathrm{Li}_{\nu+1}(-z) \).
		Applying saddle point approximation in both integrals will yield a saddle point that has a large positive real value in
		\( \alpha \) in the requested regime.
		Therefore the contour for \( z=\mathrm{e}^\alpha \) has to be thought of to be deformed outside the unit circle enclosing part of the
		branch cut.
		Nevertheless, only the behaviour in the vicinity of the saddle point shall be regarded, dropping the exact form of integration
		contours and neglecting the integration along the branch cut, allowing for an asymptotic expansion of the polylogarithm for
		large arguments \( | z | \gg 1 \) or \( \Re(\alpha) \gg 1 \).
		The way to proceed is the following. First \( \alpha \) is considered to have a large real part.
		From the consequent approximation the saddle point will be found to fulfil this statement, justifying the assumption afterwards
		in a self-consistent manner.
		The asymptotic expansion of \( \mathrm{Li}_{\nu+1} \) leaves the exponent in~(\ref{eq rho as int}) - which has the statistical
		interpretation of entropy - as
		\begin{align}
			\label{eq S assym}
			S(\beta, \alpha) \approx \beta E - \alpha N + \beta^{-\nu} \left( \frac{\alpha^{\nu+1}}{\Gamma(\nu+2)} +
				\frac{\pi^2}{6} \, \frac{\alpha^{\nu-1}}{\Gamma(\nu)} \right) \, .
		\end{align}
		The saddle point equations are
		\begin{align}
			\label{eq SPE N}
			N &= \frac{\mu^\nu}{\Gamma(\nu+1)} + \frac{\pi^2}{6 \beta^2} \, \frac{\mu^{\nu-2}}{\Gamma(\nu-1)} \, ,\\
			\label{eq SPE E}
			E &= \frac{\nu \, \mu^{\nu+1}}{\Gamma(\nu+2)} + \frac{\pi^2}{6 \beta^2} \, \frac{\nu \, \mu^{\nu-1}}{\Gamma(\nu)}
		\end{align}
		at the saddle point values of \( \beta \) and \( \mu = \alpha \slash \beta \), statistically interpreted as inverse temperature
		and chemical potential.
		The right hand sides of~(\ref{eq SPE N}) and~(\ref{eq SPE E}) match the statistical definitions of mean particle number
		\( \langle N \rangle \) and mean energy \( \langle E \rangle \) in the grand canonical formalism.
		Thus the saddle point values of \( \beta \) and \( \mu \) fix the ensemble averages of energy and number of particles
		to the given values \( \langle N \rangle = N \) and \( \langle E \rangle = E \).
		For further computation, it is convenient to express the equations in terms of the following quantities.
		\begin{equation}
			\begin{alignedat}{2}
				\varrho(\mu) &= \frac{\mu^{\nu-1}}{\Gamma(\nu)} \, , & \qquad \varrho'(\mu) &= \frac{\mu^{\nu-2}}{\Gamma(\nu-1)} \, , \\
				\mathcal{N}(\mu) &= \frac{\mu^{\nu}}{\Gamma(\nu+1)} \, , & \mathcal{E}(\mu) &= \frac{\nu \, \mu^{\nu+1}}{\Gamma(\nu+2)} \, ,
			\end{alignedat}
		\end{equation}
		with the level counting function \( \mathcal{N}(\mu) = \int_{-\infty}^\mu \dd E \, \varrho(E)  \) and the total many-body energy
		\( \mathcal{E}(\mu) = \int_{-\infty}^\mu \dd E \, E \varrho(E) \) up to \( E = \mu \).
		\( \bar{\varrho}_\spp \) has been abbreviated to \( \varrho \) in order to ease notation.
		The saddle point equations then read
		\begin{align}
			\label{eq SPE N 2}
			N &= \mathcal{N}(\mu) + \frac{\pi^2}{6 \beta^2} \, \varrho'(\mu) \, , \\
			\label{eq SPE E 2}
			E &= \mathcal{E}(\mu) + \frac{\pi^2}{6 \beta^2} \, (\mu \varrho(\mu))' \, ,
		\end{align}
		matching the Sommerfeld expansions up to the first term.
		Since \( \mu \approx E_\mathrm{F} \) at the saddle point in the requested regime, one can expand
		\begin{align}
			\label{eq N(mu)}
			\mathcal{N}(\mu) &\approx \underbrace{\mathcal{N}(E_\mathrm{F})}_{=N} + \varrho(\mu) \, (\mu - E_\mathrm{F}) \, , \\
			\Rightarrow \quad \mathcal{E}(\mu) &\approx \mathcal{E}(E_\mathrm{F}) + \mu \varrho(\mu) \, (\mu - E_\mathrm{F})
				\stackrel{(\ref{eq SPE N 2})}{\approx} E_{\text{GS}} - \frac{\pi^2}{6 \beta^2} \, \mu  \varrho'(\mu) \, ,
		\end{align}
		leading to
		\begin{align}
			\label{eq E - EGS}
			E - E_\text{GS} \approx \frac{\pi^2}{6 \beta^2} \, \varrho(\mu) \, .
		\end{align}
		The entropy at the saddle point
		\begin{align}
			S = \Bigl(1 + \frac{1}{\nu}\Bigr) \beta E - \alpha N
		\end{align}
		computes to
		\begin{align}
			\frac{1}{\beta} \, S &= (\nu+1) \frac{\mu^{\nu+1}}{\Gamma(\nu+2)} +
				(\nu + 1) \frac{\pi^2}{6 \beta^2} \, \frac{\mu^{\nu-1}}{\Gamma(\nu)} 
				- \frac{\mu^{\nu+1}}{\Gamma(\nu+1)} - \frac{\pi^2}{6 \beta^2} \, \frac{\mu^{\nu-1}}{\Gamma(\nu-1)} \, , \nonumber \\
			\Rightarrow S &= 2 \frac{\pi^2}{6 \beta^2} \, \frac{\mu^{\nu-1}}{\Gamma(\nu)}
				= 2 \frac{\pi^2}{6 \beta^2} \, \varrho(\mu) \beta = 2 \beta (E - E_\text{GS}) \, .
		\end{align}
		And by exploiting~(\ref{eq E - EGS}) one finds \( S = \sqrt{\frac{2 \pi^2}{3} \, \bar{\varrho} \, Q} \).
		What is left is the computation of the determinant of second derivatives of \( S \) with respect to \( \beta \) and \( \alpha \)
		\begin{align}
			| \det S'' | = \left| \frac{\partial( \langle N \rangle, \langle E \rangle )}{\partial ( \alpha, \beta )} \right|
				= \frac{1}{\beta} \left| \frac{\partial( \langle N \rangle, \langle E \rangle )}{\partial ( \mu, \beta )} \right| \, .
		\end{align}
		After neglecting terms that are sub-dominant in the low temperature limit like
		\( \frac{1}{\beta} \, \varrho^{(n+1)}(\mu) \ll \varrho^{(n)}(\mu) \), one finds
		\begin{align}
			| \det S'' | \approx \frac{\pi^2}{3 \beta^4} [ \varrho(\mu) ]^2 = \frac{12}{\pi^2} (E-E_\text{GS})^2 \, .
		\end{align}
		Collecting everything and accounting for a minus sign due to the directions of the integration paths in the saddle point finally gives the Bethe approximation
		\begin{align}
			\bar\varrho_-(N,E) \approx \frac{1}{\sqrt{48} \, (E-E_\text{GS})}
				\exp\left( \sqrt{ \frac{2 \pi^2}{3} \bar\varrho \cdot (E-E_\text{GS}) } \right) \, ,
		\end{align}
		with the ground state energy \( E_\text{GS} = \nu [ \Gamma(\nu+1) N ]^{1+\frac{1}{\nu}} \slash \Gamma(\nu+2) \) which matches
		the definition of smoothly counted ground state energy \( E_\text{GS}^\text{(f)} \)~(\ref{eq E GS f}).
		What is left is to show the consistency of \( \alpha \gg 1 \) at the saddle point.
		From~(\ref{eq E - EGS}) one obtains
		\begin{align}
			\alpha \approx \frac{\pi \mu \varrho(\mu)}{\sqrt{ 6 \varrho(\mu)  Q }}
				= \frac{\pi \nu \mathcal{N}(\mu)}{\sqrt{ 6 \bar\varrho \, Q }} \, .
		\end{align}
		Together with \( \mathcal{N}(\mu) = N - \frac{\pi^2}{6 \alpha^2} \nu (\nu - 1) \mathcal{N}(\mu) \) from~(\ref{eq N(mu)}),
		\(\alpha\) computes to
		\begin{align}
			\alpha \approx \frac{\pi \nu N}{\sqrt{ 6 \bar\varrho \, Q }} - \frac{\pi^2}{6} \, \frac{\nu (\nu-1)}{\alpha} \, .
		\end{align}
		Thus in the regime of low excitation energies and high numbers of particles, the assumption is self-consistently justified with
		\begin{align}
			\label{eq alpha approx}
			\alpha \approx \frac{\pi \nu N}{\sqrt{ 6 \bar\varrho \, Q }} + \mathcal{O}\Bigl( \frac{\sqrt{\bar\varrho Q}}{N} \Bigr) \, .
		\end{align}
		\begin{figure}
			\begin{center}
					\includegraphics[width=\textwidth]{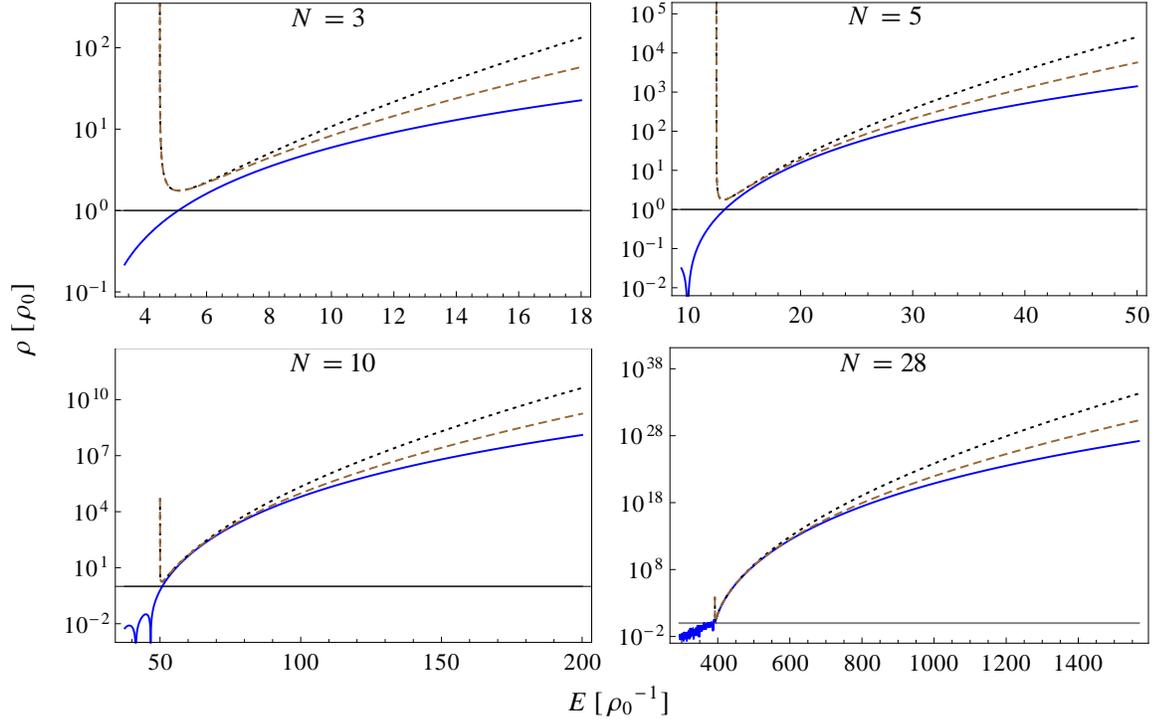}
				\caption{\label{fig:plot RvsB}Symmetry-projected DOS without boundary corrections in \( D=2 \)
				for \( N = 3, 5, 10 \) and \( 28 \) fermions (blue, solid)
				in comparison with the Bethe approximation (black, dotted) and the finite-size corrected
				improved form~(\ref{eq erdos lehner}) (brown, dashed).
				Densities are measured in units of \( \varrho_0 \)~(\ref{eq rho0}). The grey line indicates \( \varrho(E) = \varrho_0 \). }
			\end{center}
		\end{figure}%
		Figure~\ref{fig:plot RvsB} compares the Bethe approximation with the unconfined densities in two dimensions for
		different numbers of particles.
		The approximation gets better for increasing \( N \) and decreasing \( Q \) (down to \( Q \sim \varrho_0^{-1} \)).
		The displayed energy window scales with \( N^2 \).
		Thus the similar looking plots for \( N \gtrsim 10 \) show that the range of excitation energy in which
		\( \bar{\varrho}_\mathrm{B}(Q) \) gives reasonable results roughly scales also with \( N^2 \) as indicated by~(\ref{eq alpha approx}).
		Note that the Bethe approximation needs the information about the particular many-body ground state energy as additional input
		(which has here been taken as \( E_\te{GS}^\te{(f)} \)) whereas the smooth part of the unconfined symmetry-projected
		density does not but instead automatically provides it.
		In addition to \( \bar{\varrho}_\mathrm{B}(Q) \), also an improved form including finite-\( N \) corrections,
		\begin{equation}
			\label{eq erdos lehner}
			\bar{\varrho}_\text{F}(N,Q) = \bar{\varrho}_\mathrm{B}(Q) \;
				\exp\left[ \biggl( \frac{1}{2} - \frac{ \sqrt{6 \bar\varrho \, Q} }{\pi} \biggr)
				\exp \biggl( - \frac{\pi N}{ \sqrt{6 \bar\varrho \, Q} } \biggr) \right] \, ,
		\end{equation}
		is plotted. Equation (\ref{eq erdos lehner}) can be found in number theoretical context~\cite{ErdoesLehner} as approximation to
		the smooth part \( \bar{p}_N(n) \) of the number of possible partitions of an integer \( n \) where all parts are restricted
		to a maximum size \( N \).
		This function is equivalent to the non-interacting many-body DOS for an equidistant single-particle spectrum which makes the connection between number theory and many-body physics in this context.
		There is also a generalization~\cite{LeboeufFinite} of~(\ref{eq erdos lehner}) obtained by saddle point approximation in
		the statistical approach similar to the derivation of \( \bar{\varrho}_\mathrm{B}(Q) \) but including a correction
		\( \sim \mathrm{e}^{-\alpha} \) to the asymptotic expansion of the polylogarithm~(\ref{eq S assym}) in~(\ref{eq rho as int})
		for small \( \Re ( \alpha ) \).
		The resulting expressions are generalized for single-particle DOS of arbitrary power \( \bar{\varrho}_\spp(E) \sim E^{\nu-1} \).
\section{Conclusion and Outlook}
\label{sec outlook}%
	In this paper we have made explicit the essential geometrical content behind the smooth part of the density of states in
	systems of non-interacting particles.
	We have shown that both the functional form of the level density and the ground state energy are consequence of a large
	cancellation effect between contributions polynomial in the energy related with manifolds in configuration space which are
	invariant under the group of permutations. Our results cover the whole regime of energies and are valid for any number of particles.

	The main message of our work is the rigorous construction of the connection between polynomial contributions to the density
	of states and the measures and dimensions of manifolds in the classical phase space which are invariant under particular permutations of the particles.
	Moreover, by means of algebraic manipulations we have transformed our expression in a way that explicitly shows its equivalence
	with several known but previously unconnected results.
	In particular, by re-expressing the sum over cluster zones in terms of generating functions, we have established its equivalence
	with the thermodynamic formalism leading to the celebrated Bethe estimate for the level density in fermionic systems.

	The geometry of invariant manifolds presented here is a purely kinematic concept independent on any particular dynamics,
	and although we have used it to fully construct the level density for non-interacting systems~(\ref{eq rho sym unconfined 2},\ref{eq rho sym confined}), we expect that it plays also a
	fundamental role when interparticle interactions are present.
	In order to apply our formalism in this more general scenario, the key step is to relax the condition expressed by~(\ref{eq K non-int}) which makes explicit the assumption of non-interacting particles.
	In the case of interacting particles, however, the key concept remains the same:
	the smooth part of the density of states does not require the solution of the problem including both interactions and
	confinement simultaneously.
	Symmetry effects on the Weyl expansion for interacting systems are encoded in unconfined (but interacting) propagation
	around cluster zones.

	In order to keep the problem tractable, well controlled approximations for the unconfined (but still interacting)
	many-body propagator are necessary, and the first task to extend our methods into interacting systems is to check the physical
	picture one gets when a certain approximation for the unconfined propagator is used around the cluster zones.
	One possible scenario where this program can be carried on is when we treat the interactions in eikonal approximation,
	freezing the interactions between sets of particles in different cluster zones. This and other possible approximations will
	be reported in a future publication.

\ack
	The authors thank Matthias Brack for carefully reading the paper.

\appendix
\setcounter{section}{0}

\section{Calculation via Convolution Formula - unconfined case}
\label{app calc conv unconfined}

    The single-particle Weyl volume term reads
    \begin{align}
      \bar{\varrho}_\spp(E) = \left( \frac{m}{2 \pi \hbar^2} \right)^\frac{D}{2} \; V_D \;
				\frac{E^{\frac{D}{2}-1}}{\Gamma\left(\frac{D}{2}\right)} \, \theta(E)
				\equiv a \; E^{\frac{D}{2}-1} \, \theta(E) \, .
    \end{align}
		Using~(\ref{eq weidenmuller conv}) yields the sum
		\begin{align}
			\bar{\varrho}_{\te{conv},\pm}(E) = \frac{1}{N!} \sum_{l=1}^N (\pm 1)^{N-l} \sum_{\substack{N_1,\ldots,N_l =1\\
			N_1 \leq \cdots \leq N_l}}^N \delta_{N, \, \sum N_\omega} \;\; c(N_1,\ldots,N_l)
				\Bigl( \prod_{\omega=1}^l \frac{1}{N_\omega} \Bigr) \; \mathcal{C}(E)
    \end{align}
    of convolutions \( \mathcal{C}(E) \) of the form
    \begin{align}
			\label{eq C(E)}
      \mathcal{C}(E) = &\int \dd E_1 \ldots \dd E_l
				\, \left[ \prod_{\omega=1}^l \bar{\varrho}_\spp\Bigl( \frac{E_\omega}{N_\omega} \Bigr) \right] \,
				\delta \Bigl( E-\sum_{\omega=1}^l E_\omega \Bigr) = a^l  \frac{\partial}{\partial E} \mathcal{I}(E) \, , \\
			\mathcal{I}(E) := &\Bigl( \prod_{\omega=1}^l \frac{1}{N_\omega} \Bigr)^{\frac{D}{2}-1} \;
				\int_0^\infty \dd E_1 \ldots \dd E_l \;
				\prod_{\omega=1}^l E_\omega^{\frac{D}{2}-1} \;
				\theta \Bigl( E-\sum_{\omega=1}^l E_\omega \Bigr) \, .
    \end{align}
    \( \mathcal{I}(E) \) can be calculated by successively performing the integrals of single-particle energies
		\( E_\omega \) and solving an emerging recursion relation.
    It is convenient to write \( r=D \slash 2 -1 \) and define a new variable for the energy to be distributed among the first \( n \)
    particles for every \( n \)
    \begin{align}
      e_n = E - \sum_{\omega=n+1}^l E_\omega , \qquad n=0,\ldots,l-1 \, .
    \end{align}
    The first step in the integral of \( \mathcal{I}(E) \) is then
    \begin{align}
				\int_0^\infty \dd E_1 E_1^r \theta(e_1 - E_1) &= \frac{1}{r+1} e_1^{r+1} \theta(e_1)
				= \frac{1}{r+1} (e_2 -E_2)^{r+1} \theta(e_2 - E_2) \, .
    \end{align}
    The next integral over \( E_2 \) is of the form
    \begin{align}
      \label{eq recursion integral}
      \int_0^c \dd x \; x^r \, (c-x)^s \theta(c)= c^{s+r+1} \frac{\Gamma(1+r) \Gamma(1+s)}{\Gamma(2+r+s)} \theta(c) \, ,
    \end{align}
    with \( c = e_2 = e_3 - E_3 \). Therefore, also the third integral and all others are of this form. Let
    \( A_n \) be the total prefactor and \( s_n \) the exponent \( s \) in~(\ref{eq recursion integral}) appearing
    in the \( n \)-th integral step. Then one gets the recurrence relation
    \begin{equation}
      \begin{alignedat}{2}
				A_{n+1} &= A_n \frac{\Gamma(1+r) \Gamma(1+s_n)}{\Gamma(2+r+s_n)} \, , & \quad A_1 &= 1 \, , \\
				s_{n+1} &= r + s_n + 1 \, , & s_1 &= 0 \, .
      \end{alignedat}
    \end{equation}
		The convolution~(\ref{eq C(E)}) is then expressed as
		\begin{align}
      \mathcal{C}(E) = \Bigl( \prod_{\omega=1}^l \frac{1}{N_\omega} \Bigr)^{\frac{D}{2}-1} \, a^l \, A_{l+1} \, s_{l+1} \,
				E^{s_{l+1}-1} \theta(E) \, .
    \end{align}
    The solution of the recurrence reinserting \( r=D \slash 2 - 1 \) is
    \begin{align}
      A_{n+1} = \frac{\left[ \Gamma\left( \frac{D}{2} \right) \right]^n}{ \Gamma\left( \frac{n D}{2} + 1 \right) } \, , \quad
      s_{n+1} = \frac{n D}{2} \, , \qquad n \geq 0 \, .
    \end{align}
    Using the the original expression for \( a \) one gets the final result
    \begin{align}
      \begin{split}
      \bar{\varrho}_{\te{conv},\pm}(E) = &\frac{1}{N!} \sum_{l=1}^N (\pm 1)^{N-l} \sum_{\substack{N_1,\ldots,N_l =1\\ N_1 \leq \cdots \leq N_l}}^N
				\delta_{N, \, \sum N_\omega} \;\; c(N_1,\ldots,N_l) \\
			&\times \; \left( \frac{m}{2 \pi \hbar^2} \right)^\frac{l D}{2}
				\Bigl( \prod_{\omega=1}^l \frac{1}{N_\omega} \Bigr)^\frac{D}{2} \; V_D^l \;
				\frac{E^{\frac{l D}{2}-1}}{\Gamma\left(\frac{l D}{2}\right)} \, \theta(E) \, ,
      \end{split}
    \end{align}
		which is identical to the result obtained by investigating the propagation in cluster zones~(\ref{eq rho sym unconfined}).
    The usage of the volume term in the single-particle Weyl expansion is consistent with neglecting the single-particle
    billiard boundaries in the calculation of short path propagations.
\section{Convolution Property of the Confined Propagator}
\label{app K conv prop confined}
	Let the confined propagator for a single particle in first correction be denoted by
	\begin{align}
		K({\bf q}_2, {\bf q}_1; t) = K_0({\bf q}_2, {\bf q}_1; t) \pm K_0(R{\bf q}_2, {\bf q}_1; t) \, ,
	\end{align}
	where \( \pm \) refers to van Neumann or Dirichlet-boundary conditions respectively.
	\( R{\bf q}_2 \) denotes thereby the reflection of the point \( {\bf q}_2 \) with respect to the boundary,
	locally regarded as being flat.
	In the spacial integral that has finally to be performed to obtain the DOS, for points lying far inside the
	billiard compared to the wavelength in question the second term won't contribute.
	This justifies regarding \( R \) as the unambiguous reflection with respect to the tangent plane in the nearest boundary point.
	Thus, whenever a point far inside the billiard is formally reflected in the expressions below, one should keep in
	mind that then the expression won't contribute and that therefore any ambiguity does no harm.
	The assumption of flatness also implies
	\begin{align}
		\label{eq K reflect}
		K_0({\bf q}', {\bf q}; t) = K_0(R{\bf q}', R {\bf q}; t) \, ,
	\end{align}
	where \( {\bf q}', {\bf q} \) are arbitrary positions not restricted to the interior of the billiard.
	The convolution of two propagators where all of the three involved positions (initial, final and intermediate position) are
	lying in the interior \( \Omega \) of the billiard reads
	\begin{align}
		\int\limits_\Omega \dd^D q_2 \, K({\bf q}_3, {\bf q}_2; t') K({\bf q}_2, {\bf q}_1; t)
			= \; &\int\limits_\Omega \dd^D q_2 \, K_0({\bf q}_3, {\bf q}_2; t') K_0({\bf q}_2, {\bf q}_1; t) \nonumber\\
				+  &\int\limits_\Omega \dd^D q_2 \, K_0(R{\bf q}_3, {\bf q}_2; t') K_0(R{\bf q}_2, {\bf q}_1; t) \nonumber\\
			  \pm &\int\limits_\Omega \dd^D q_2 \, K_0(R{\bf q}_3, {\bf q}_2; t') K_0({\bf q}_2, {\bf q}_1; t) \nonumber\\
			\label{eq K conf conv}
			 \pm &\int\limits_\Omega \dd^D q_2 \, K_0({\bf q}_3, {\bf q}_2; t') K_0(R{\bf q}_2, {\bf q}_1; t) \, .
	\end{align}
	Using~(\ref{eq K reflect}) and \( R^2={\rm id} \) turns~(\ref{eq K conf conv}) into
	\begin{align}
		&\int\limits_\Omega \dd^D q_2 \, K_0({\bf q}_3, {\bf q}_2; t') K_0({\bf q}_2, {\bf q}_1; t)
				+  \int\limits_\Omega \dd^D q_2 \, K_0({\bf q}_3, R{\bf q}_2; t') K_0(R{\bf q}_2, {\bf q}_1; t) \nonumber\\
			 \pm &\int\limits_\Omega \dd^D q_2 \, K_0(R{\bf q}_3, {\bf q}_2; t') K_0({\bf q}_2, {\bf q}_1; t)
			 \pm \int\limits_\Omega \dd^D q_2 \, K_0(R{\bf q}_3, R{\bf q}_2; t') K_0(R{\bf q}_2, {\bf q}_1; t) \nonumber \\
		= &\int\limits_{\mathclap{\Omega \cup R(\Omega)}} \dd^D q_2 \, K_0({\bf q}_3, {\bf q}_2; t') K_0({\bf q}_2, {\bf q}_1; t)
			\pm \int\limits_{\mathclap{\Omega \cup R(\Omega)}} \dd^D q_2 \, K_0(R{\bf q}_3, {\bf q}_2; t') K_0({\bf q}_2, {\bf q}_1; t) \, .
	\end{align}
	As the intermediate coordinates \( {\bf q}_2 \) now are no longer restricted to the interior \( \Omega \) but instead can
	be regarded as running over full space, the convolution property of the free propagator can be applied to obtain
	\begin{align}
		\int\limits_\Omega \dd^D q_2 \, K({\bf q}_3, {\bf q}_2; t') K({\bf q}_2, {\bf q}_1; t)
			&= K_0({\bf q}_3, {\bf q}_1; t + t') \pm K_0(R{\bf q}_3, {\bf q}_1; t + t') \nonumber \\
		&= K({\bf q}_3, {\bf q}_1; t+t') \, .
	\end{align}
	This shows that the assumption of local flatness of the boundary suffices to give the confined propagator the convolution
	property needed for the convolution formula by Weidenmüller to hold also when using it in combination with the single-particle
	Weyl expansion up to the first boundary correction instead of the exact DOS.
\section{Calculation via Convolution Formula - confined case}
\label{app calc conv confined}
	As in the unconfined case, the many-body DOS can be computed from convolutions of single-particle DOS in Weyl expansion.
	The single-particle Weyl expansion including the first boundary correction reads
    \begin{align}
      \bar{\varrho}_\spp(E) &= \varrho_0 \frac{(\varrho_0 E)^{\frac{D}{2}-1}}{\Gamma\left( \frac{D}{2} \right)} \theta(E)
				+ \varrho_0 \gamma \frac{(\varrho_0 E)^{\frac{D}{2}-\frac{3}{2}}}{\Gamma\left( \frac{D}{2} - \frac{1}{2} \right)} \theta(E) \nonumber \\
				&\equiv a E^{\frac{D}{2}-1} \theta(E) + b E^{\frac{D}{2}-\frac{3}{2}} \theta(E) \, .
    \end{align}
		Using~(\ref{eq weidenmuller conv}) and writing the sum of partitions as ordered tuples yields the sum
		\begin{align}
			\label{eq rho conv C}
			\bar{\varrho}_{\te{conv},\pm}(E) = \sum_{l=1}^N \frac{(\pm 1)^{N-l}}{l!} \sum_{\substack{N_1,\ldots,N_l =1 \\ \sum N_\omega = N}}^N
				\Bigl( \prod_{\omega=1}^l \frac{1}{N_\omega} \Bigr)^2 \; \mathcal{C}(E)
    \end{align}
    of convolutions \( \mathcal{C}(E) \) of the form
		\begin{align}
			\mathcal{C}(E) &= \int \dd E_1 \ldots \dd E_l \, \left[ \prod_{\omega=1}^l \bar{\varrho}_\spp
				\Bigl( \frac{E_\omega}{N_\omega} \Bigr) \right]
				\; \delta\Bigl( E - \sum_{\omega=1}^l E_\omega \Bigr) \nonumber \\
			\label{eq C(E) conf}
			&= \sum_{l_{\rm V} = 0}^l \genfrac(){0pt}{}{l}{l_{\rm V}} a^{l_{\rm V}} b^{l-l_{\rm V}} \frac{\partial}{\partial E} \mathcal{I}(E) \, ,
		\end{align}
		where
		\begin{align}
			\mathcal{I}(E) = \Bigl( \prod_{\omega=1}^{l_\text{V}} \frac{1}{N_\omega} \Bigr)^{\frac{D}{2}-1}
				\Bigl( \prod_{\omega=l_\text{V}+1}^l \frac{1}{N_\omega} \Bigr)^{\frac{D}{2}-\frac{3}{2}} \tilde{\mathcal{I}}(E) \, ,
		\end{align}
		with
		\begin{align}
			\tilde{\mathcal{I}}(E) = \int_0^\infty \dd E_1 \ldots \dd E_l
				\Bigl( \prod_{\omega = 1}^{l_{\rm V}} E_\omega^{\frac{D}{2}-1} \Bigr)
				\Bigl( \prod_{\omega = l_{\rm V}+1}^l E_\omega^{\frac{D}{2}-\frac{3}{2}} \Bigr) \theta\Bigl( E - \sum_{\omega=1}^l E_\omega \Bigr) \, .
		\end{align}
		The reason for expressing~(\ref{eq rho conv C}) as the sum over ordered tuples is the resulting invariance of the sum with respect
		to relabelling amongst the \( N_1, \ldots, N_l \) in the summands. This allows to subsume all summands in
		\( \prod_{\omega=1}^l \bar{\varrho}_\spp(E_\omega \slash N_\omega) \) with certain powers in \( a \) and \( b \).
		The number of such equivalent summands is counted by the binomial coefficient in~(\ref{eq C(E) conf}).
		Again the partial sums of energies
		\begin{align}
      e_n = E - \sum_{\omega=n+1}^l E_\omega , \qquad n=0,\ldots,l
		\end{align}
		are defined.
		The first \( l_{\rm V} \) integrals then are identical to the unconfined case with the substitutions \( l \rightarrow l_{\rm V} \)
		and \( E \rightarrow e_{l_{\rm V}} \), yielding
		\begin{align}
			\tilde{\mathcal{I}}(E) &= \frac{\Gamma\left( \frac{l_{\rm V} D}{2} + 1 \right)}{\Gamma\left( \frac{D}{2} \right)^{l_{\rm V}}}
				\int_0^\infty \dd E_{l_{\rm V}+1} \ldots \dd E_l
				\left. \prod_{\omega = l_{\rm V}+1}^l E_\omega^{\frac{D}{2}-\frac{3}{2}} \right.
				( e_{l_{\rm V}} )^\frac{l_{\rm V} D}{2} \theta ( e_{l_{\rm V}} ) \, .
		\end{align}
		The \( l_{\rm S} = l - l_{\rm V} \) remaining integrals are all again of the form~(\ref{eq recursion integral}) where this time
		\( r = \frac{D}{2} - \frac{3}{2} \). Also the initial coefficient \( A_1 \) and exponent \( s_1 \) are different, all in all leading to
		the following recurrence relation
		\begin{equation}
			\begin{alignedat}{2}
				A_{n+1} & = A_n \frac{\Gamma(1+r) \Gamma(1+s_n)}{\Gamma(2+r+s_n)} \, , & \quad A_1 & = \frac{\Gamma\left( \frac{l_{\rm V} D}{2} + 1 \right)}{\Gamma\left( \frac{D}{2} \right)^{l_{\rm V}}} \, , \\
				s_{n+1} & = r + s_n + 1 \, , & s_1 & = \frac{l_{\rm V} D}{2} \, ,
			\end{alignedat}
		\end{equation}
		solved by
		\begin{align}
			\begin{split}
				s_{n+1} &= s_1 + n ( r + 1) \, , \\
				A_{n+1} &= A_1 \Gamma(1+r)^n \frac{\Gamma(1+s_1)}{\Gamma(1+ s_{n+1})} \, .
			\end{split}
		\end{align}
		Recognizing that
		\begin{align}
			\tilde{\mathcal{I}}(E) = A_{l_{\rm S}+1} E^{s_{l_{\rm S}+1}} \theta(E) \, ,
		\end{align}
		and reintroducing the expressions for \( a \) and \( b \) leads to
		\begin{align}
			\begin{split}
				\mathcal{C}(E) = &\varrho_0 \sum_{l_{\rm V} = 0}^l \genfrac(){0pt}{}{l}{l_{\rm V}}
					\Bigl( \prod_{\omega=1}^{l_\text{V}} \frac{1}{N_\omega} \Bigr)^{\frac{D}{2}-1}
					\Bigl( \prod_{\omega=l_\text{V}+1}^l \frac{1}{N_\omega} \Bigr)^{\frac{D}{2}-\frac{3}{2}} \\
				&\times \gamma^{l_\text{S}} \frac{( \varrho_0 E)^{\frac{l_\text{V} D}{2} + \frac{l_\text{S} (D-1)}{2} - 1}}
					{\Gamma\bigl( \frac{l_\text{V} D}{2} + \frac{l_\text{S} (D-1)}{2} \bigr)}
					\theta(E) \, ,
			\end{split}
		\end{align}
		which, inserted into~(\ref{eq rho conv C}), gives exactly the expression~(\ref{eq rho sym confined}).

		One has to mention that for \( D=1 \) the second term in the single-particle Weyl expansion has to be replaced
		according to
		\begin{align}
			\label{eq D=1 rule SP}
			\frac{ (\varrho_0 E)^{\frac{D-1}{2} -1 }}{ \Gamma \left( \frac{D-1}{2} \right) } \theta(E)
				\stackrel{D\rightarrow1}{\longrightarrow} \delta(\varrho_0 E) \, .
		\end{align}
		Nevertheless this does not heavily change the obtained result, as in the case \( l_{\rm V} > 0 \) the integrals involved
		instead of~(\ref{eq recursion integral}) are of the form
		\begin{align}
			\int_{-\infty}^c \dd x \, \delta(x) (c-x)^s = c^s \theta(c)
				= \left. c^{s+r+1} \frac{\Gamma(1+s)}{\Gamma(2+r+s)} \theta(c) \right|_{r=-1} \, ,
		\end{align}
		which leads to a similar recurrence relation. Recognizing that the missing factor of \( \Gamma(1+r) \) in the end is just compensated
		by the replacement of \( \Gamma((D-1) \slash 2) \) in~(\ref{eq D=1 rule SP}) this shows the validity of all terms with
		\( l_{\rm V} > 0 \) also in the case \( D=1 \).
		The summand corresponding to \( l_{\rm V} = 0 \) involves convolutions of Dirac-Delta distributions only, leading to the
		already mentioned replacement rule~(\ref{eq D=1 rule MB}) in the final expression similar to~(\ref{eq D=1 rule SP}).

		It has also to be mentioned that analogous to the presented calculation it is possible to include also higher terms
		in the single-particle Weyl expansion (\( D \geq 2 \)) corresponding to corrections originating in the curvature of the boundary.
		The resulting expressions are then involving an additional summation over an index \( l_{\rm C} \) representing the number of
		clusters contributing via curvature corrected propagation. The already time intense computation of the coefficients \( C_{l, l_{\rm V}} \)
		(\ref{eq univ coeff confined}) without curvature corrections then becomes even worse in the further extended case.
		Also the property~(\ref{eq conv property K}) had then to be shown to link the smooth part of the DOS with the convolution formula
		of Weidenmüller.

		As examples (see section~\ref{sec geom emergence}) show, the effect of the curvature in singly connected billiards generically is not
		as strong as the effect of the surface correction and might be approximated by simply shifting the energy corresponding to
		the associated \( E_\text{GS}^\text{(f)} \). Therefore and because of the analogous and straightforward but extensive derivation of the
		corresponding DOS, this computation is not explicitly shown here.\\
\\

\bibliographystyle{my-iopart-num}
\bibliography{lit2}

\end{document}